\pdfoutput=1
\documentclass{aa}  

\usepackage{graphicx}
\usepackage{textcomp}
\usepackage{txfonts}
\usepackage{natbib,twoopt}
\usepackage[breaklinks=true]{hyperref} 
\bibpunct{(}{)}{;}{a}{}{,} 
\makeatletter
\newcommandtwoopt{\citeads}[3][][]{\href{http://adsabs.harvard.edu/abs/#3}%
{\def\hyper@linkstart##1##2{}%
\let\hyper@linkend\@empty\citealp[#1][#2]{#3}}}
\newcommandtwoopt{\citepads}[3][][]{\href{http://adsabs.harvard.edu/abs/#3}%
{\def\hyper@linkstart##1##2{}%
\let\hyper@linkend\@empty\citep[#1][#2]{#3}}}
\newcommandtwoopt{\citetads}[3][][]{\href{http://adsabs.harvard.edu/abs/#3}%
{\def\hyper@linkstart##1##2{}%
\let\hyper@linkend\@empty\citet[#1][#2]{#3}}}
\newcommandtwoopt{\citeyearads}[3][][]%
{\href{http://adsabs.harvard.edu/abs/#3}
{\def\hyper@linkstart##1##2{}%
\let\hyper@linkend\@empty\citeyear[#1][#2]{#3}}}
\makeatother

\usepackage{mathabx}

\begin{document}

   \title{Expected performances of the Characterising Exoplanet Satellite (CHEOPS)}
   \subtitle{II. The CHEOPS simulator}

   \author{D.~Futyan\inst{\ref{unige}}\and A.~Fortier\inst{\ref{unibe}}\and M.~Beck\inst{\ref{unige}}\and D.~Ehrenreich\inst{\ref{unige}}\and A.~Bekkelien\inst{\ref{unige}}\and W.~Benz\inst{\ref{unibe}}\and N.~Billot\inst{\ref{unige}}\and V.~Bourrier\inst{\ref{unige}}\and C.~Broeg\inst{\ref{unibe}}\and A.~Collier~Cameron\inst{\ref{standrews}}\and A.~Deline\inst{\ref{unige}}\and T.~Kuntzer\inst{\ref{unige}}\and M.~Lendl\inst{\ref{unige},\ref{graz}}\and D.~Queloz\inst{\ref{unige},\ref{cambridge}}\and R.~Rohlfs\inst{\ref{unige}}\and A.~E.~Simon\inst{\ref{unibe}}\and F.~Wildi\inst{\ref{unige}}}

   \institute{Département D'Astronomie, Université De Genève, Chemin des Maillettes 51, 1290 Versoix, Switzerland\\\email{david.futyan@unige.ch}\label{unige}\and
     Physikalisches Institut, University of Bern, Sidlerstrasse 5, 3012 Bern, Switzerland\label{unibe}\and
     School of Physics and Astronomy, Physical Science Building, North Haugh, St Andrews, United Kingdom\label{standrews}\and
     Space Research Institute, Austrian Academy of Sciences, Schmiedlstr. 6, 8042 Graz, Austria\label{graz}\and
     Battcock Centre for Experimental Astrophysics, Cavendish Laboratory, JJ Thomson Avenue, Cambridge CB3 0HE, United Kingdom\label{cambridge}
             }

   \date{Received September 2, 2019; accepted January 13, 2020}

 
  \abstract
   {The CHaracterising ExOPlanet Satellite (CHEOPS) is a mission dedicated to the search
for exoplanetary transits through high precision photometry of bright stars already known to host
planets. The telescope will provide the unique capability of determining accurate radii for planets
whose masses have already been measured from ground-based spectroscopic surveys. This will allow a first-order characterisation of the planets' internal structure through the determination of the bulk density,
providing direct insight into their composition.
By identifying transiting exoplanets with high potential for in-depth characterisation, CHEOPS will also provide prime
targets for future instruments suited to the spectroscopic characterisation of exoplanetary
atmospheres.}
   {The CHEOPS simulator has been developed to perform detailed simulations of the data which is to be received from the CHEOPS satellite.
It generates accurately simulated images that can be used to explore  design options and to test the on-ground data processing, in particular, the pipeline producing the photometric time series. It is, thus, a critical tool for estimating the photometric performance expected in flight and to guide photometric analysis. It can be used to prepare observations, consolidate the noise budget, and asses the performance of CHEOPS in realistic astrophysical fields that are difficult to reproduce in the laboratory.}
   {The simulator has been implemented as a highly configurable tool called CHEOPSim, with a web-based user interface. Images generated by CHEOPSim take account of many detailed effects, including variations of the incident signal flux and backgrounds, and detailed modelling of the satellite orbit, pointing jitter and telescope optics, as well as the CCD response, noise and readout.}
   {The simulator results presented in this paper have been used in the context of validating the data reduction processing chain, in which image time series generated by CHEOPSim were used to generate light curves for simulated planetary transits across real and simulated targets. Independent analysts were successfully able to detect the planets and measure their radii to an accuracy within the science requirements of the mission: For an Earth-sized planet with an orbital period of 50 days orbiting a Sun-like target with magnitude $V$=6, the median measured value of the planet to star radius ratio, $R_p/R_s$, was 0.00923 $\pm$ 0.00054(stat) $\pm$ 0.00019(syst), compared to a true input value of 0.00916. For a Neptune-sized planet with an orbital period of 13 days orbiting a target with spectral type K5V and magnitude $V$=12, the median measured value of $R_p/R_s$ was 0.05038 $\pm$ 0.00061(stat) $\pm$ 0.00031(syst), compared to a true input value of 0.05.}
   {}


   \maketitle
%

\section{Introduction}

In recent years, the number of exoplanet discoveries has rapidly increased, with over four thousand planets confirmed as of June 2019 \citep{NASA_Exoplanet_Archive}\footnote{
According to the NASA Exoplanet Archive, which can be accessed at https://exoplanetarchive.ipac.caltech.edu/}. Many of these, including Earth-sized planets, primarily orbiting faint stars, have been discovered using the transit technique by the {\it CoRoT}~\citep{CoRoT} and {\it Kepler}~\citep{kepler} space missions, while others, mainly giants, primarily orbiting bright stars, have been discovered through ground-based radial velocity surveys such as HARPS~\citep{HARPS} and HARPS-N~\citep{HARPS-N}. However, there are few exoplanets with masses lower than around 30 Earth masses that have precise measurements of both mass and radius due to the fact that the CoRoT and Kepler targets are too faint for the Doppler measurement with current spectroscopic facilities.

The search for rocky planets amenable to atmospheric characterisation (i.e. transiting bright stars) has led the Community and space agencies to adopt new dedicated survey missions: the {\it Transiting Exoplanet Survey Satellite} (TESS, NASA)~\citep{TESS} and PLATO~\citep{PLATO}, as well as a fast-track follow-up mission: the {\it CHaracterising ExOPlanet Satellite} (CHEOPS)~\citep{CHEOPS}, a 'small mission' jointly developed by ESA and a consortium of eleven European countries led by Switzerland.

{
Although other missions such as the Hubble Space Telescope (HST) and {\it Spitzer}~\citep{Spitzer} have conducted follow up observations of previously detected exoplanets, CHEOPS is the first mission dedicated to the precise measurement of exoplanet radii through the use of high precision photometric time series observations of bright stars ($V$-band magnitude < 12) already known to host planets.} The telescope will have access to more than 70\% of the sky and it will provide the unique 
capability of determining accurate radii of planets whose masses have already been measured from ground-based spectroscopic
surveys, with sufficient precision to detect Earth-sized transits. This will allow a first-order characterisation of the planets' internal structure through the determination of the bulk density,
providing direct insight into their composition.
CHEOPS will also provide precise radii for new exoplanets of Neptune size and smaller discovered by the next generation of
ground- or space-based transits surveys, in particular by TESS, which was launched in April 2018 with the purpose of performing
a whole-sky transit search survey. CHEOPS will be able to observe targets identified by TESS for longer periods and with the benefit of a larger telescope aperture, allowing more precise radius measurements, as well as the potential to discover additional planets in the system. By identifying transiting
exoplanets with high potential for in-depth characterisation, CHEOPS will also provide prime
targets for future instruments suited to the spectroscopic characterisation of exoplanetary
atmospheres such as JWST~\citep{JWST} and E-ELT~\citep{E-ELT}.

The CHEOPS satellite carries a single instrument which consists of an optical Ritchey-Chrétien telescope with 30~cm effective aperture diameter and a large external baffle to minimise stray light, which is critical due to the low Earth orbit. The telescope delivers a defocussed image of the target star onto a single frame-transfer CCD detector\footnote{
The CCD is a back illuminated sensor manufactured by Teledyne e2V, part number CCD47-20, featuring Advanced Inverted Mode Operation (AIMO) to minimise dark current. For details see https://www.e2v.com/resources/account/download-datasheet/1427} on the focal plane covering the wavelength range 330–1100~nm with a field of view of 0.32 degrees. The tube assembly is passively cooled and thermally controlled to support high precision, low noise photometry. The point spread function (PSF) has a 90\% encircled energy radius of approximately 12 pixels, to ensure the illumination of sufficient pixels to minimise the photometric effect of satellite tracking residuals. The CCD and the
front-end electronics are both thermally stabilised at the precision
of 10 mK, with operating temperatures of -40\textdegree C and -10\textdegree C
respectively, in order to limit the noise contributions of the dark
current and electronic gain variability. In order to maximise science data down-link and to achieve a one minute sample rate, a circular cut out of a 200$\times$200 pixel sub-frame 
centered on the target is downloaded. It represents
a field of view of 3.3 arcminutes in diameter. The pixel scale on the detector
corresponds to 1.002 arcseconds on the sky.

The spacecraft will orbit the Earth in a polar Sun Synchronous Orbit (SSO) above the terminator at an altitude of 700~km and a local time of the ascending node (LTAN) of 6:00~am. The orbit inclination is about 98\textdegree\ and the orbital period is just under 100 minutes. The performance of the payload attitude and control system (AOCS) is designed to maintain tracking on target to better than 4 arcseconds (rms). During the course of its orbit around the Earth, the spacecraft continuously rolls so that its cold plate radiators always face away from the Earth. As a result, the field of view rotates around the pointing direction.

The nominal mission duration is 3.5 years, with a goal to five years. {
CHEOPS was successfully launched on 18th December 2019.}

In order to meet its scientific objectives, CHEOPS has been designed to measure photometric signals with an ultra-high precision of 20 ppm in six hours of integration time for a 9th magnitude star and 85 ppm in three hours of integration for a 12th magnitude star, with ultra-high stability being maintained over at least 48 hours, with an observation cadence better than one minute. These requirements are challenging due to the combination of a number of noise sources, some of which involve complex interactions between the components of the instrument which cannot be modelled analytically. For example, as a result of spacecraft pointing jitter, the position of the PSF moves with time relative to the CCD pixel grid. {
Although the photometric aperture follows the PSF centroid, the motion results in photometric variations due to non-uniformities in the pixel-to-pixel response, adding noise to the light curve.} Another effect is that of field of view rotation when there is a bright background star close to the target, which, due to the irregular shape of the point spread function, can result in periodic modulations to the light curve. The CHEOPS simulator has been developed to address and precisely quantify aspects such as these through detailed simulation of the spacecraft motion, the telescope optics, and the CCD response.

\begin{figure*}
\centering{
\resizebox{.7\hsize}{!}{\includegraphics{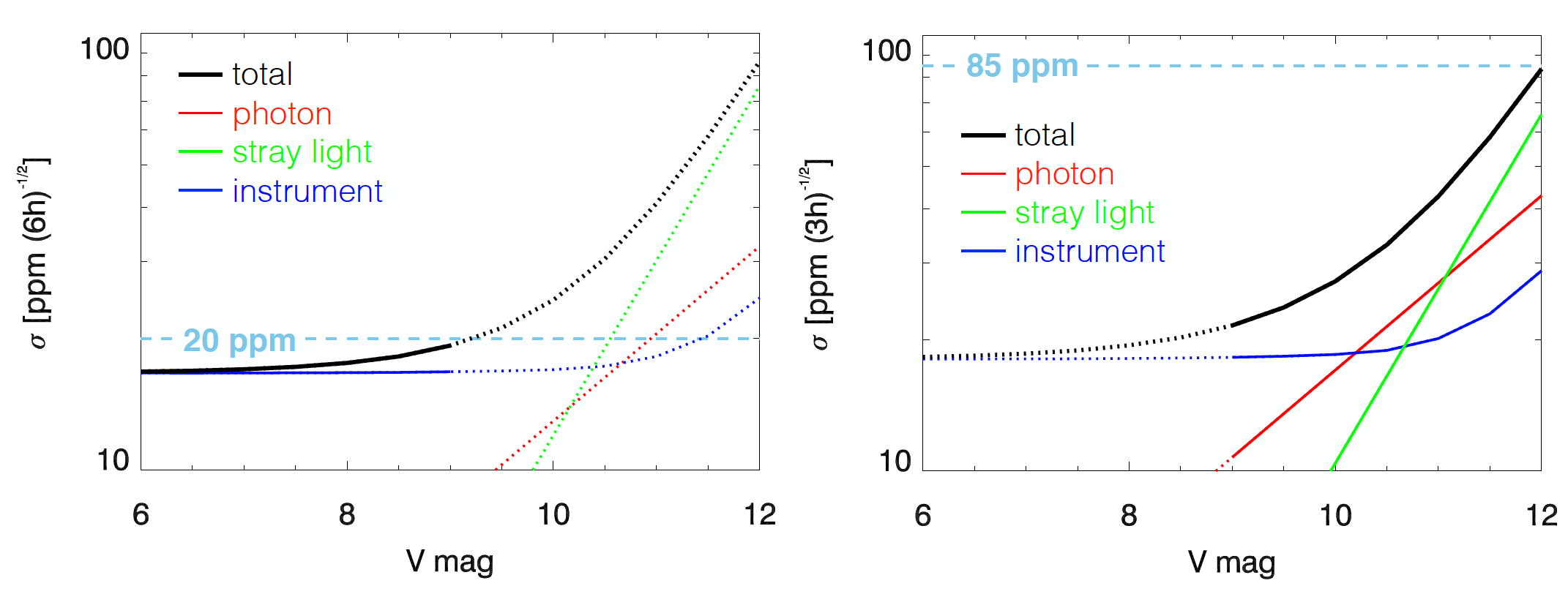}}
}
\caption{Expected photometric performance of CHEOPS. (Left panel) For bright stars (here assumed to be G-type stars with an
effective temperature of 5500 K and apparent magnitude $6\leq V<9$), the rms of the light curve residuals is required to be less than 20~ppm
in six hours of integration, limited mainly by instrumental noise. (Right panel) For faint stars (here assumed to be K-type
stars with an effective temperature of 4500 K and apparent magnitude $9\leq V\leq12$), the rms of the light curve residuals
is required to be less than 85~ppm in three hours of integration, dominated by instrument noise for $9\leq V<10$, by photon noise for $10\leq V<11$ and
by stray light for $V\geq 11$. {
We note that unlike photon noise, instrument noise does not scale quadratically with the integration time.}}
\label{fig:noise_budget}
\end{figure*}

The overall noise budget {
of the CHEOPS instrument} can be broken down into three main components: photon noise of the incident light from the target and background sources, instrumental noise resulting from jitter, CCD response and readout, and noise due to stray light contamination. The contribution of these three components to the expected photometric performance of CHEOPS is shown as a function of stellar $V$ magnitude in Fig.~\ref{fig:noise_budget}. {

Although stellar variability is modelled by CHEOPSim (see Sect.~\ref{sec:stellar_variation}), it is not considered part of the CHEOPS noise budget, as it depends on the target and is independent of the instrument performance. In the case of the {\it Kepler} mission, \citet{Gilliland1} and \citet{Gilliland2} found the contribution to noise from stellar variability to be around 20 ppm in six hours.}
{
A difference with the Kepler sample however is that CHEOPS targets are bright ($6 \leq V \leq 12$) and already known to host exoplanets. Knowing the planet periods from existing radial velocity measurements, for instance, will help retrieving planetary signals in photometric time series that will be eventually limited by stellar variability. Activity indices retrieved from high-resolution spectroscopy can be used to estimate the stellar rotation period (e.g. the calcium index; \citet{Mamajek_Hillenbrand}), also helping with the disambiguation of stellar signals. However, the stellar noise linked with stellar p-modes (granulation and super-granulation) will remain a strong limiting factor to the real photometric precision obtained for a given star.}

This paper is part of a mini-series, {\it Expected performances of the Characterising Exoplanet Satellite (CHEOPS)}, which consists of three papers. In addition to this paper, the series includes {\it Photometric performances from ground-based calibration} \citep{Deline}, which describes the on-ground calibration of the detector, the results of which are used as input to the simulator described here, and presents the results of processing and photometric analysis of images recorded during the calibration, as well as images from the simulator. The series also includes {\it The CHEOPS data reduction pipeline: architecture and performances} \citep{Hoyer}, which describes the data reduction pipeline used to generate the light curves presented here, in which raw images are processed to correct for instrumental response effects, such as flat field and CCD non-linearity, and environmental effects such as background stars, cosmic rays, and field of view rotation; and in which light curves are generated through photometric extraction.

\section{Model}
\subsection{Overview}

{
End-to-end science data simulators have been used with great success for earlier space missions including Kepler \citep{Bryson,Jenkins1} and TESS \citep{Jenkins2,Smith}.}

The CHEOPS simulator, called CHEOPSim, is highly configurable software capable of performing detailed simulations of the data which will be received from the CHEOPS satellite, including image time series, associated metadata, and housekeeping data, all using the same format, file structures, and file naming as will be used for real data.

The simulator has two main use cases: firstly, it provides data that can be used as input to facilitate the development and testing of the on-ground data processing chain at the Software Operations Centre (SOC). The processing chain includes the pre-processing of the RAW data received at the Misson Operations Centre (MOC), Quick Look software for fast inspection of the data, and the data reduction processing chain \citep{Hoyer}. For the purposes of testing the pre-processing chain, unstacked images generated by CHEOPSim are passed through a separate piece of software, the Data Flow Simulator, which executes the on-board software to compress the images into the bitstream format that will be sent to ground from the spacecraft.

The second use case is to provide simulated images and light curves to provide more detailed understanding of the capability of CHEOPS to observe transits for potential targets.

The images generated by CHEOPSim take account of many detailed effects including variations of the incident signal flux, due to stellar fluctuations and variability as well as planetary transits (Sect.~\ref{sec:flux}); modelling of the satellite orbit and pointing jitter, including field of view rotation, and interruptions due to Earth occultation and the South Atlantic Anomaly (Sect.~\ref{sec:satellite}); an interface with the Gaia star catalogue in order to generate a field of view, projected onto the plane of the CCD (Sect.~\ref{sec:fieldofview}); modelling of the telescope optics, including the point spread function, scattering halo and ghosts  (Sect.~\ref{sec:telescope}); modelling of the background from zodiacal and stray light (Sect.~\ref{sec:background}); and detailed modelling of the CCD response, noise and readout, including flat field, dark current, shot noise, bad pixels, cosmic rays, full well saturation, frame transfer, charge transfer efficiency, gain non-linearity and read noise (Sect.~\ref{sec:detector}).

The output of CHEOPSim, and the web interface used to configure and execute the simulations are described in Appendices~\ref{app:output} and \ref{app:web}, respectively.

\subsection{Modelling of the incident signal flux}
\label{sec:flux}

For each star in the field of view, the {
photon flux spectrum as a function of wavelength $\lambda$} is calculated given the the magnitude and spectral type either read from the Gaia catalogue \citep{gaia}, or input manually by the user. An effective temperature is assigned to the spectral type according to \citet{Mamajek}.
{
For effective temperatures between 2300K to 7200K (corresponding to spectral types M9V to F0V), the energy spectrum is read from a library of spectral energy distributions (SEDs), generated using the PHOENIX model \citep{PHOENIX}. For higher effective temperatures, the energy spectrum is approximated by a Planck black body. The photon spectrum is calculated as the energy spectrum multiplied by the energy per photon, $hc/\lambda$.}

The spectrum is first normalised to a magnitude of 0.035 by requiring that the integral of the spectrum in the $V$-band is equal to the integrated flux of Vega in the $V$-band. The flux is then renormalised according to the value of the input magnitude $m$ by multiplying by 10$^{m-0.035}$. These calculations are performed with a wavelength resolution of 1~nm over the range 330~nm to 1100~nm.

The variation of the incident flux with time for the target star is modelled as described below in Sects.~\ref{sec:transit} and \ref{sec:stellar_variation}. As an alternative to using these models, the user has the possibility to provide an externally generated incident flux time series as an ascii file.

\subsubsection{Transit model}
\label{sec:transit}

The transit curve is calculated as described by \citet{MandelAgol}.  Without limb darkening, the transit curve is entirely defined by a two parameters: the time independent parameter $p$, defined as the ratio of the planet radius to the star radius, and the time dependent parameter $z(t)$, defined as the separation between the star and the planet centres divided by stellar radius, $R_\star$.  These two parameters are derived from user input values for the transit mid-time, the radius and orbit period of the planet, and the transit impact parameter, together with the mass and radius of the star, which are assigned depending on the input spectral type according to \citet{Mamajek}, with $z(t)$ calculated as:

\begin{equation}
z(t)=\sqrt{\left(\frac{a}{R_\star}\sin(2\pi f(t))\right)^2 + (b\cos(2\pi f(t)))^2},
\label{eq:transit1}
\end{equation}
where $b$ is the impact parameter, $f(t)$ is the difference between the current time $t$ and the transit midpoint time as a fraction of the orbit period, and $a$ is the semi-major axis of the planets' orbit, given by $a=\sqrt[3]{G M_\star/\left(2\pi/P\right)^2}$, where $M_\star$ is the mass of the star and $P$ is the orbit period.

Given $p$ and $z(t)$, the multiplication factor to be applied to the flux as a function of time is calculated using the formulae in Sect. 2 of \citet{MandelAgol}.

Limb darkening is implemented according to the equations for quadratic limb darkening in Sect. 4 of \citet{MandelAgol}. The formulae require two coefficients, whose values are calculated depending on the spectral type using the algorithm described in \citet{limbDarkening}, using software provided by the authors, and using the ATLAS model \citep{claret_bloeman} to perform the fit. The algorithm takes as input the effective temperature and surface gravity of the star, assigned for a given spectral type according to \citet{Mamajek}, and a response function, defined according to the telescope optical throughput and the CCD quantum efficiency. The stellar metallicity is assumed to be zero, and the microturbulent velocity to be 2~km/s.

\subsubsection{Stellar granulation and variability}
\label{sec:stellar_variation}

CHEOPSim takes as input a set of 48 hour time series, each containing deviations from the nominal stellar flux due to granulation in steps of 15 seconds. Such time series have been generated for several values of the mass, radius and effective temperature of the star (Lendl et. al., in prep.). {
The time series were calculated based on representative power spectra, in which the component of granulation was modelled by a Harvey \citep{Harvey} function, which had been slightly adapted as suggested by e.g. \citet{Kallinger}. The granulation time scale and amplitude were extrapolated from the Sun to other main-sequence stars using the relations of \citet{Gilliland2}, and the results were validated against Kepler data presented by \citet{Cranmer}.}

CHEOPSim also simulates the flux modulation produced by stellar active regions (spots and plages) as they rotate in and out of view over the stellar rotation period. This is modelled using a Gaussian process in time $t$, with a quasi-periodic kernel function (see Eq. 5.16 in \citet{variation}) that produces a covariance with a periodic behaviour modulated by a decay away from exact periodicity:
\begin{equation}
k(t_i,t_j) = A\cdot \mathrm{exp}\left(-\frac{(t_i-t_j)^2}{2\tau^2} - \frac{2\sin^2(\pi (t_i-t_j)/P)}{\mu^2}\right) ,
\end{equation}
where the four required parameters are the amplitude $A$ (in mmag), the decay time $\tau$, the stellar rotation period $P$, and the smoothing or structure parameter $\mu$.
The stellar rotation period is provided by the user, the amplitude and decay time are taken randomly from the sample generated in \citet{HelenGiles}, based on the spectral type of the target star, and the dimensionless structure parameter $\mu$ is randomly drawn between 0.5 and 1.0.

{
Shorter time scale variations caused by flares are not taken into account by this model. However, since multiple transit observations are planned for most targets, the effects of such temporally isolated events can be mitigated, since they will not be present in all observed transits for a given target.} 

\begin{figure}
\resizebox{\hsize}{!}{\includegraphics{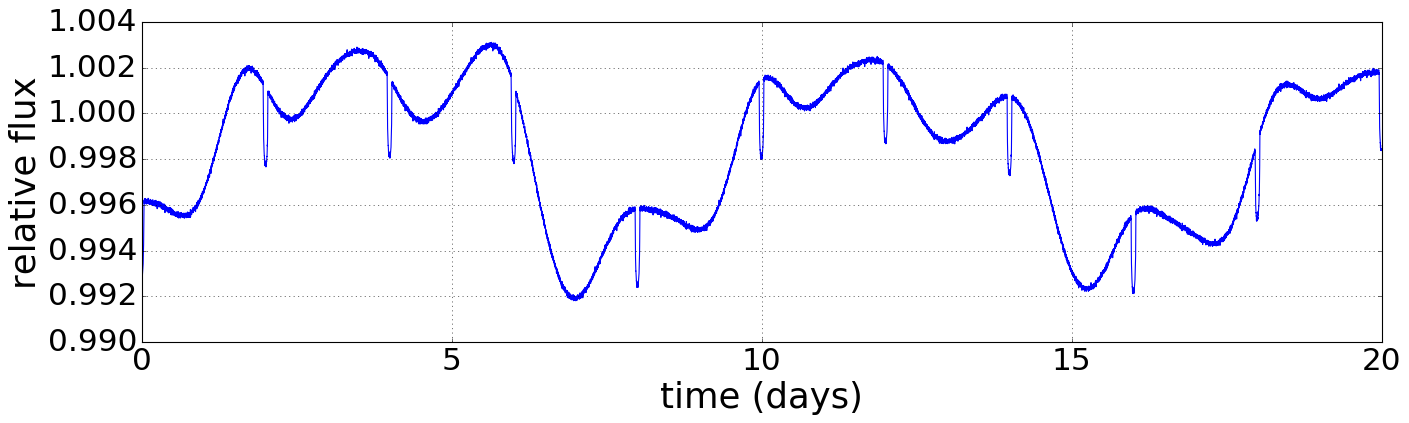}}
\caption{Example incident flux time series over a period of 20 days, including stellar granulation (barely visible), stellar variability due to spots and plages, and transits, for a Neptune-sized planet with an orbit period of 2 days orbiting a K4 star with a rotation period of 8.3 days.}
\label{fig:variation}
\end{figure}

Figure~\ref{fig:variation} shows an example incident flux time series over a period of 20 days, including granulation, variability due to spots and plages, and transits.

\subsection{Modelling of the satellite orbit and pointing}
\label{sec:satellite}

\subsubsection{Orbit model}
\label{sec:orbit}

We use typical orbit trajectory data specifying the position of the spacecraft in Earth Centred Inertial frame coordinates (EME2000) for a Sun synchronous orbit at an altitude of 700~km, for Local Time of Ascending Node (LTAN) {
6:00~am}, calculated at one minute intervals between 31 Dec 2018 00:00 and 2 Aug 2022 00:00. Recall that the spacecraft rolls during its orbit to keep its cold plate radiators facing away from the Earth, resulting in a rotation of the field of view around the pointing direction. The roll angle at any given time is calculated using as input the pointing direction, the position vector and the velocity vector of the spacecraft in the intertial frame at that time. Rotation matrices are derived to transform these quantities between the inertial frame, the Local Vertical / Local Horizontal (LVLH) frame, the orbital frame, and the satellite frame. The roll angle $\phi$ is then extracted from the matrix elements of the 3$\times$3 rotation matrix from the inertial to the satellite frame, $M_\mathrm{i2sat}$, as follows:
\begin{equation}
\phi = \mathrm{atan2}( M_\mathrm{i2sat}[1,2], M_\mathrm{i2sat}[2,2] ),
\end{equation}
where $M_\mathrm{i2sat}[i,j]$ indicates the matrix element of $M_\mathrm{i2sat}$ at row $i$, column $j$, with $i$ and $j$ starting from 0. {
The implementation has been validated against calculations performed by ESA.}

The position of the satellite in its orbit at any given time is used to determine the timing of interruptions due to Earth occultation, during which images are discarded, and due to the South Atlantic Anomaly, during which images are either discarded or have an enhanced cosmic ray flux (Sect.~\ref{sec:cosmics}), according to the user configuration.

\subsubsection{Pointing jitter}
\label{sec:jitter}

As described in the introduction, spacecraft pointing jitter is an important noise source, since it results in the point spread function moving over the CCD pixel grid, leading to photometric variations due to non-uniformities in the pixel-to-pixel response (Sect.~\ref{sec:flat}). It is therefore important that both jitter and pixel-to-pixel response variations are accurately modelled.\footnote{
CHEOPSim simulations have shown that the contribution to noise in the light curve from this effect is below 3~ppm in six hours integration time, which is lower than the requirement of 5~ppm.}

\begin{figure}
\resizebox{\hsize}{!}{\includegraphics{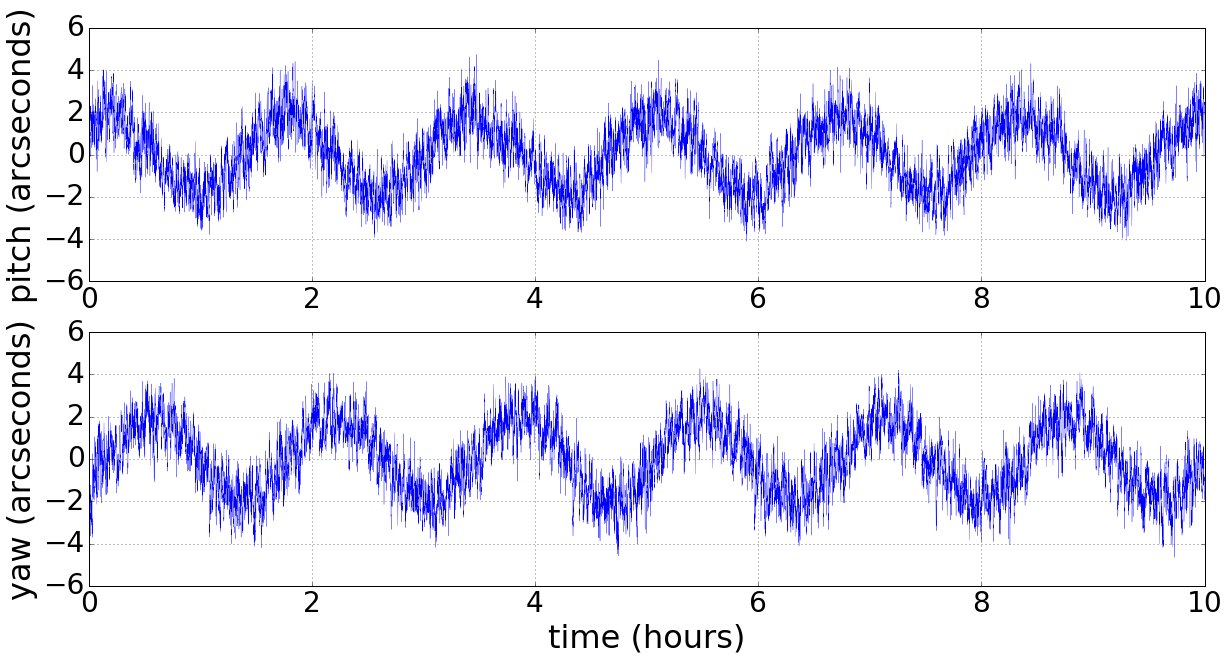}}
\caption{First ten hours of the simulated time series for the Y and Z components (corresponding to the pitch and yaw of the spacecraft, respectively) of the absolute pointing errors for spacecraft jitter, for the nominal case of four reaction wheels, two star trackers, 60 second centroid sampling cadence, with no interruptions, and with the payload always in the loop. The sinusoidal pattern is a result of thermo-elastic deformation errors between centroid measurements. The combination of the Y and Z component oscillations is a circular motion of the pointing direction with period corresponding to the orbit period.}
\label{fig:jitter}
\end{figure}

Spacecraft attitude jitter is modelled using a pre-calculated time series of Absolute Pointing Errors (APEs) in three dimensions at one second intervals over a 48 hour period. Several such time series have been provided, with different configurations for the number of star trackers, the number of reaction wheels, the centroid sampling cadence, and whether or not the payload is always in the loop. For the nominal configuration, the 2D pointing error is 2.4 arcseconds at 68\%. The first ten hours of the time series for the nominal case is illustrated in Fig.~\ref{fig:jitter}.

The APEs are used to calculate the perturbations to the pointing direction and to the roll angle at any given time. The perturbation to the pointing direction is calculated as follows:

\begin{equation}
m_\mathrm{point} = {M_\mathrm{i2sat}}^T\ dR\ M_\mathrm{i2sat}\ m_\mathrm{nominalPoint},
\end{equation}

where $m_\mathrm{nominalPoint}$ and $m_\mathrm{point}$ are the pointing vectors in cartesian coordinates before and after applying the APEs, and dR is the APE rotation matrix:

\begin{equation}
dR = \begin{bmatrix} 
1 & -APE_Z & APE_Y \\ 
APE_Z & 1 & -APE_X \\ 
-APE_Y & APE_X & 1  
\end{bmatrix}.
\end{equation}

The perturbation to the roll angle is equal to the X APE component.

\subsection{Simulation of the field of view}
\label{sec:fieldofview}

A list of stars within the field of view can be provided by the user either directly through the web interface or uploaded as a file in ascii or fits format. A tool is provided to extract a list of stars from the Gaia star catalogue in the required fits format.

For each second of the exposure, the star positions are projected onto the focal plane of the detector, taking into account the jitter and field of view rotation defined as described in Sect.~\ref{sec:satellite}, using a World Coordinate System routine included within the CFITSIO software library~\citep{cfitsio}, using a gnomonic (tangent plane) projection. The projection algorithm takes as input: the right ascension and declination of each star in the field of view; the right ascension and declination of the jittered pointing direction and the jittered roll angle for each second of the exposure, calculated as described in Sect.~\ref{sec:satellite}; the intended location of the target on the CCD, corresponding to the axis of rotation of the field of view; and the plate scale of the CCD (1.002 arcsecond per pixel).

The dimensions and position offset of the image sub-array (by default the central 200$\times$200 pixels, corresponding to 3.33$\times$3.33 arcminutes on the sky), as well as the intended target location on the CCD, can be configured by the user.

\subsection{Modelling of telescope optics}
\label{sec:telescope}

\subsubsection{Point spread function}
\label{sec:psf}

\begin{figure}
\resizebox{\hsize}{!}{\includegraphics{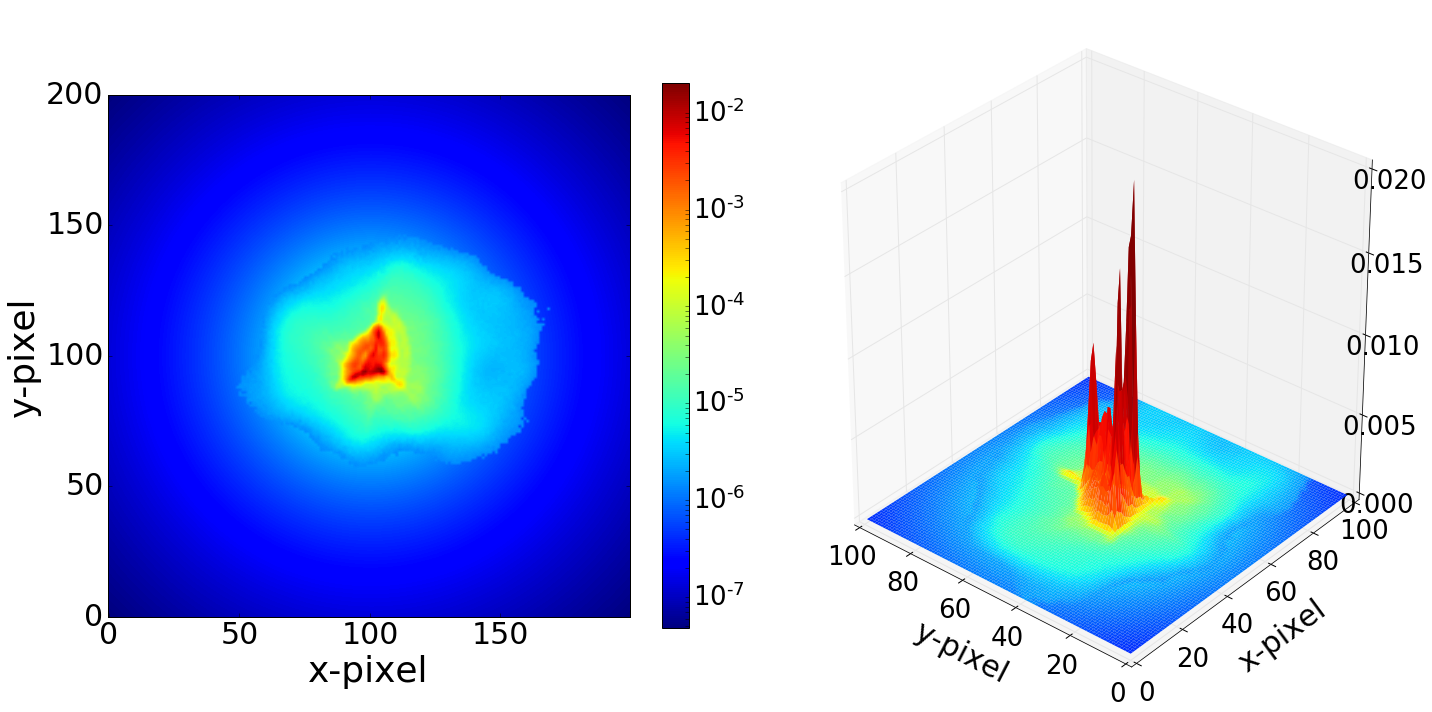}}
\caption{Point spread function from laboratory measurements performed during the CHEOPS on-ground calibration. Despite smoothing with a bilateral filter, the tails of the PSF measurement are affected by noise, and have therefore been replaced with a $1/r^{3}$ function for pixels below a threshold value. The image is normalised such that the sum over all pixels is 1. The field corresponds to 3.33$\times$3.33 arcminutes on the sky. The right plot shows a zoom on the central 100$\times$100 pixels in 3D, illustrating more clearly the presence of spikes.}
\label{fig:psf}
\end{figure}

A point spread function (PSF) is generated at the position of each star on the focal plane (see Sect.~\ref{sec:fieldofview}). An empirical PSF from laboratory measurements performed during the CHEOPS on-ground calibration is used by default, shown in Fig.~\ref{fig:psf}. The PSF can be seen to contain sharp spikes, which enhance the effect of pixel-to-pixel response non-uniformities as the PSF moves over the CCD pixel grid due to jitter (see Sect.~\ref{sec:jitter}).

In addition to the empirical PSF, various synthetic PSF models are available in CHEOPSim in order to investigate certain effects: PSFs modelled for a set of different temperatures of the telescope optics allow to investigate the effect of telescope breathing, and PSFs modelled for a set of different wavelengths allow to investigate the dependence of the PSF shape on the spectrum of the incident flux and hence on the spectral type of the target. Synthetic PSFs are also available for various misalignments of the optical system, and for different positions on the CCD.

\begin{figure}
\resizebox{\hsize}{!}{\includegraphics{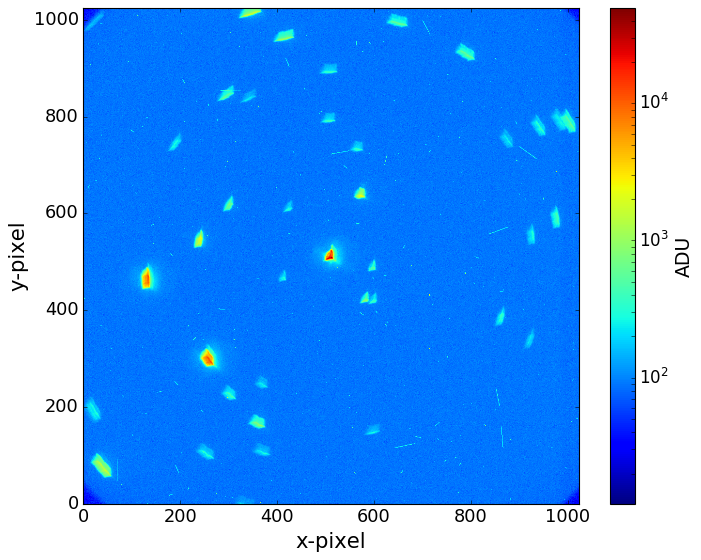}}
\caption{Sample full frame image, corresponding to the entire 1024$\times$1024 pixel CCD, for an exposure duration of 60 seconds. Smearing of the background star PSFs is visible as a result of field of view rotation during the exposure. The field corresponds to 17.1$\times$17.1 arcminutes on the sky. In general full frame images such as this are only downloaded once at the start of a visit, with 200$\times$200 pixel subframes centred on the intended target location being downloaded for subsequent exposures.}
\label{fig:fov_rotation}
\end{figure}

For a given star, the integral of the PSF is normalised to the incident flux from the star calculated as described in Sect.~\ref{sec:flux}, integrated over the telescope aperture, taken to be an unobstructed circle with diameter 32~cm. Given the position of the star at any given point in time, the 200$\times$200 pixel PSF image is overlaid onto the CCD pixel grid with the centre of the PSF grid positioned at the location of the star. The flux assigned to each CCD pixel is calculated by bilinear interpolation, using the fluxes in the four nearest pixels of the overlaid PSF, weighted according to the distance of the centres of those pixels from the centre of the current CCD pixel. The tails of the diffractive PSF beyond the 200$\times$200 grid are modelled using a $1/r^3$ function. This process is performed for all stars in the field of view, separately for each second of the exposure, as the star positions move due to pointing jitter and field of view rotation as described in Sect.~\ref{sec:fieldofview}. The images generated for each second of the exposure are summed. This results in a smearing of the PSF over the exposure, and for sufficiently long exposures, arcs being generated for off axis stars due to the field of view rotation. An example of a full frame image, corresponding to the entire 1024$\times$1024 pixel CCD, for a 60 second exposure is shown in Fig.~\ref{fig:fov_rotation}.

\subsubsection{Scattering and ghosts}

In addition to the PSF due to diffraction described in Sect.~\ref{sec:psf}, there is additional flux from scattered light due to surface microroughness and particulate dust, and from ghosts due to internal reflections. Together, these form an extended halo around the diffractive PSF. The scattering halo is modelled for each star in the field of view using a simplified version of the Peterson Model~\citep{Peterson}. Using Eq. 20 of Peterson with parameter values corresponding to the CHEOPS telescope, for an on-axis star with incident flux 1 photon/mm$^2$ (before optical throughput), the flux on the detector due to scatter at a radial distance $R$ from the centre of the CCD is given by:
\begin{equation}
E(R)=395\times[1+19R^2]^{-1.135}.
\end{equation}
For a given star, the number of photons incident on a given pixel due to scatter is given by the incident photon flux from the star (Sect.~\ref{sec:flux}) multiplied by the area of one pixel, multiplied by $E(R)$, where $R$ is the distance in mm from mean position of the star, averaged over jitter during the exposure, to the centre of the pixel. The contribution from scattering dominates over that from diffraction for $R$ greater than 155 pixels.

The total ghost flux for a given star is defined to be a fraction of the flux from the star incident on the telescope aperture. The value of that fraction varies between 0.007\% and 0.06\%, depending on the angular distance of the star from the line of sight. The flux is assumed to be uniformly distributed over the 1024$\times$1024 pixels of the exposed part of the CCD. For a given star, the total flux from ghosts integrated over the CCD is factor 11.5 lower than that from scatter. 

\subsection{Modelling of background light sources}
\label{sec:background}

\subsubsection{Zodiacal light}
\label{sec:zodiac}

The zodiacal light flux as a function of wavelength, $f_{ZL}(\lambda) $, is taken from \citet{HST_ETC}. The wavelength integrated flux corresponding to a $V$-band surface brightness of 22.1 mag arcsec$^{-2}$, is calculated as:

\begin{equation}
\label{eq:zodiacal_vs_wavelength}
F_\mathrm{\ zodiacal\ light}^{\ \mathrm{mag}=22.1} = A_\mathrm{telescope}\int_{3300\AA}^{11000\AA}{\frac{f_{ZL}(\lambda)}{hc/\lambda}d\lambda},
\end{equation}

where $A_\mathrm{telescope}$ is the telescope collecting area (unobstructed circle with diameter 32~cm). The integral evaluates to 7.857 photons s$^{-1}$ arcsecond$^{-1}$.

The brightness of the zodiacal sky background varies as a function of the angular separation between the pointing direction of the telescope and the position of the Sun in ecliptic polar coordinates. This variation is provided in terms of  $V$-band magnitude per square arcsecond in Table 9.4 of \citet{HST_ETC}. The zodiacal light magnitude $m_\mathrm{ZL}$ for arbitrary angular separations between the pointing direction and the Sun are calculated by bilinear interpolation of the values in the table. This directional dependence is taken into account by multiplying the flux from Eq.~\ref{eq:zodiacal_vs_wavelength} by 10$^{m_\mathrm{ZL}-22.1}$.

The photon flux from zodiacal light is added to the image uniformly across the exposed part of the CCD, using the plate scale of 1.002 arcseconds per pixel.

\subsubsection{Stray light}
\label{sec:straylight}

For faint stars ($V\gtrsim11$), the noise budget of CHEOPS is expected to be dominated by stray light contamination, particularly from the Earth. CHEOPS must therefore avoid pointing in directions which yield high stray light. The primary sources of stray light are the Sun, illuminated Earth limb and the Moon. The angular separations between the Sun and the line of sight, and between the Moon and the line of sight, are required to exceed exclusion angles of 120\textdegree\ and 5\textdegree, respectively. Stray light from the Earth limb undergoes strong variation as a function of the satellite orbit, with some stray light reaching the telescope aperture for angles between the line of sight and Earth limb up to approaching 90\textdegree.

The mission planning software determines the optimal scheduling sequence for CHEOPS observations. It takes into account constraints due to stray light, Earth occultation and the South Atlantic Anomaly, and calculates the stray light arriving at the CCD from sunlight reflected from the surface of the Earth as a function of time \citep{stray_light}. The distribution of the intensity of reflected light over the Earth's surface is calculated at any given time according to the position of the Sun and the angle of incidence of the sunlight on the Earth's surface. The position of the satellite relative to this distribution is then used to calculate the photon flux arriving at the telescope aperture. The Point Source Transmission function (PST) of the telescope is then used to calculate the fraction of the photons incident on the aperture that reach the CCD, given the angle of incidence.

Observations are planned so that the stray light contamination does not represent more than 5~ppm for bright stars (where we aim to reach a precision of 20~ppm) and 70~ppm for faint stars (where we aim a precision of 85~ppm). These thresholds are defined assuming that the background correction is able to correct for 99.6\% of the stray light contamination.

CHEOPSim takes as input either the stray light time series calculated by the scheduling software for the visit to be simulated as described above, or a time series provided by the user in an ascii file. The photon flux from stray light is added to the image, uniformly distributed across the exposed part of the CCD.

\subsection{Optical throughput and quantum efficiency}

The initial image generated as described in Sects. \ref{sec:fieldofview}, \ref{sec:telescope} and \ref{sec:background} corresponds to the incident photon flux (without photon noise). We describe here the conversion from the number of incident photons to the number of photoelectrons generated in each pixel (without noise).

Each incoming photon $p$ of a given wavelength $\lambda$ has a certain probability $P[p\rightarrow e]$ to result in the generation of a signal electron $e$ in the pixel on which it is incident, corresponding to the telescope optical throughput $OT$ at the wavelength of the photon multiplied by the quantum efficiency of the CCD at temperature $T_\mathrm{CCD}$ and at the wavelength of the photon:

\begin{equation}
P[p\rightarrow e](\lambda,T_\mathrm{CCD}) = OT(\lambda) \times QE(\lambda,T_\mathrm{CCD}).
\label{eq:electronProbability1}
\end{equation}

The number of electrons $N_\mathrm{e} $ in each pixel is calculated as the number of incident photons $N_\mathrm{p}$ multiplied by the integral over wavelength of the product of $P[p\rightarrow e]$ and the {
stellar photon spectrum $S$ of the target star (see Sect.~\ref{sec:flux}) normalised to 1}:

\begin{equation}
N_\mathrm{e} = N_\mathrm{p} \int_{\lambda} S(T_\mathrm{eff},\lambda) \times P[p\rightarrow e](\lambda,T_\mathrm{CCD}),
\label{eq:electronProbability2}
\end{equation}

where $T_\mathrm{eff}$ is the effective temperature of the target star\footnote{To be fully correct, the integral should be performed separately for photons from each source in the field of view, since each star will have a different $T_\mathrm{eff}$. However, for reasons of computational efficiency, $T_\mathrm{eff}$ is assumed to be the same as that of the target star for all sources.}. The integral is calculated over the wavelength range 330~nm and 1100~nm with resolution of 0.5~nm.

\begin{figure}[htbp]
\resizebox{\hsize}{!}{\includegraphics{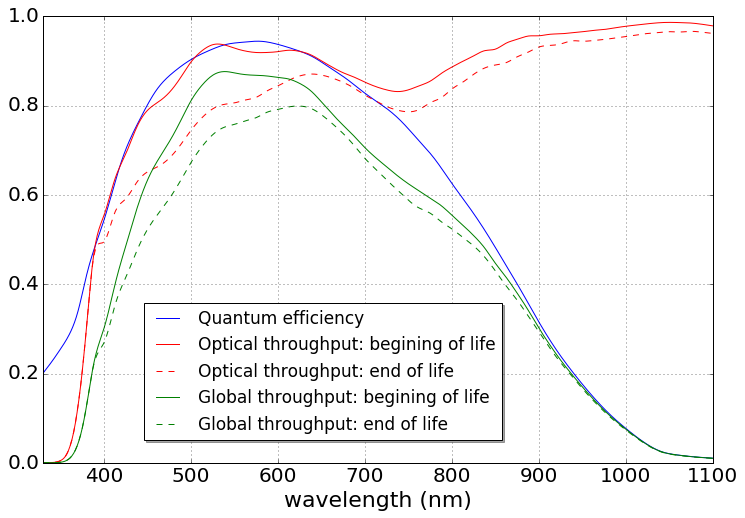}}
\caption{Telescope optical throughput and CCD quantum efficiency as a function of wavelength. The optical throughput at end of life takes into account the degradation of the optical performance of each element of the CHEOPS optical chain due to the expected radiation environment. {
The global throughput, defined as the product of the optical throughput and the quantum efficiency is also shown.}}
\label{fig:throughput_qe}
\end{figure}

The optical throughput at the beginning and end of the mission, including the effect of obscuration due to the secondary mirror and spiders, calculated using Zemax\textregistered\  simulation software, and the quantum efficiency for the CCD, {
measured by ESA\footnote{
The measurement has been verified by e2v.}}, are shown as a function of wavelength in Fig.~\ref{fig:throughput_qe}.

\subsection{Modelling of the CCD response, noise and readout}
\label{sec:detector}

\subsubsection{Flat field}
\label{sec:flat}

The importance of accurate modelling of the pixel-to-pixel response non-uniformity, which is described by the flat field, was highlighted in Sect.~\ref{sec:jitter}.

\begin{figure}[htbp]
\resizebox{\hsize}{!}{\includegraphics{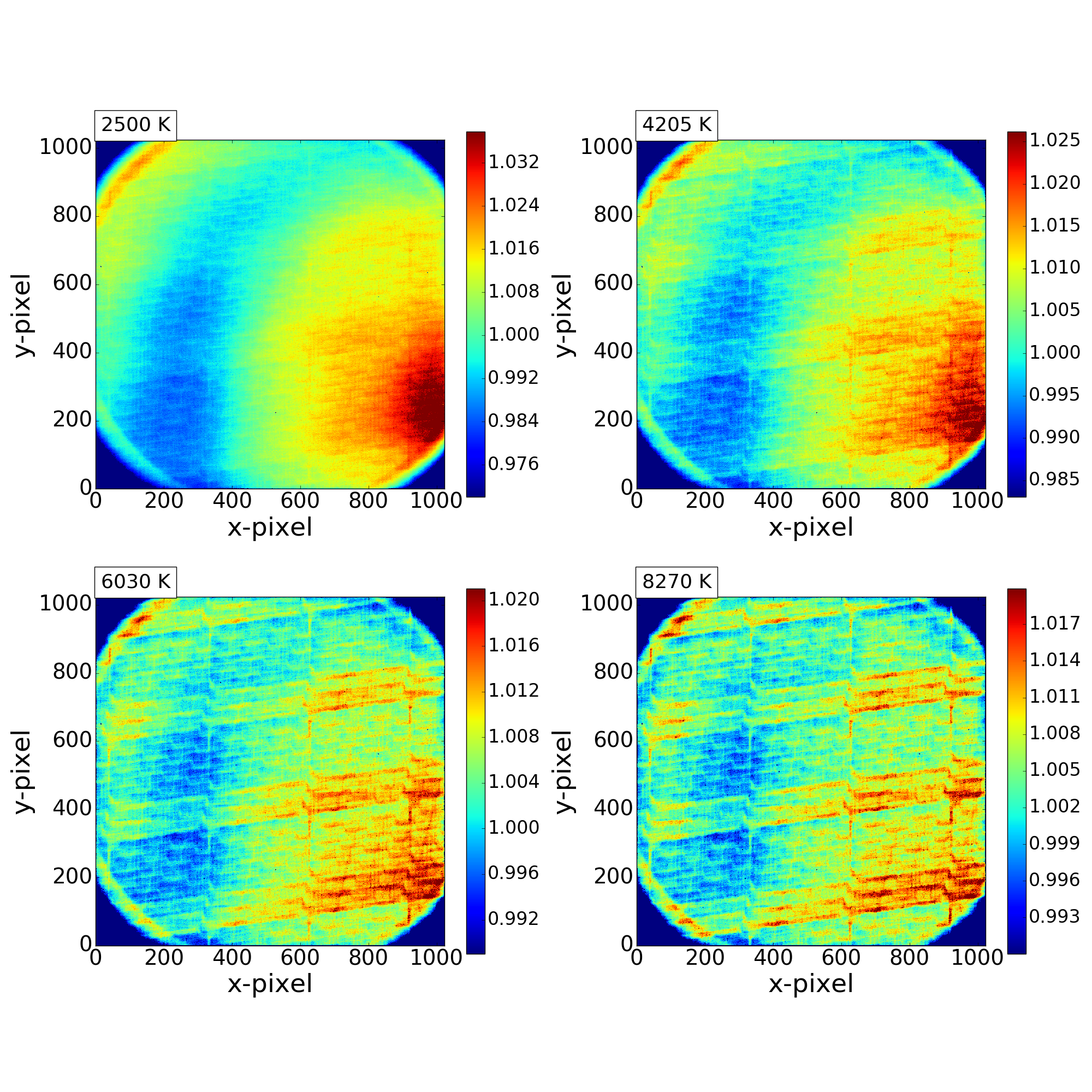}}
\caption{Empirical flat field frames weighted according to the effective temperature of the target star, for four example effective temperatures. The corners appear to have low response as a result of partial illumination.}
\label{fig:flatfield}
\end{figure}

A set of empirical flat field frames were obtained from laboratory measurements as part of the on-ground calibration of the CCD, performed using monochromatic and broad band filters for a range of wavelengths \citep{Deline}. The flat field applied to CHEOPSim images depends on the effective temperature of the target star: it is constructed by combining the laboratory measurements, weighting them as a function of wavelength according to the product of the optical throughput, quantum efficiency and the stellar spectrum of the target star. Figure~\ref{fig:flatfield} shows the flat field corresponding to four example target star effective temperatures. The flat field can optionally be modified, either through Gaussian smearing, or by applying a shift to the effective temperature, so that the flat field applied in CHEOPSim is not identical to that used to correct the flat field in data reduction (which is otherwise calculated in the same way), thereby simulating uncertainties in the reference flat field.

\subsubsection{Dark current}

Dark current arises due to the generation of thermal electrons in addition to signal photoelectrons, and is thus an additional source of random noise in the images and therefore the light curves.

An empirical dark frame was obtained from laboratory measurements as part of the on-ground calibration of the CCD \citep{Deline}. It is constructed as the average of several exposures taken at -40\textdegree C, after subtracting the bias and dividing by the gain and by the exposure duration, thus providing a dark frame with units of electrons per second per pixel. {
The mean pixel value of the resulting dark frame is 0.056 electrons per second.} The dark frame is multiplied by the accumulation time for the exposure and added to the CHEOPSim image, representing the expected number of dark electrons per pixel (before shot noise). The accumulation time is defined as the repetition period, plus the read out delay for the pixel, which depends on its location on the CCD due to the sequence in which pixels are read out, and also depends on the readout mode, which determines whether all pixels or only a subset are read out.

\subsubsection{Bad pixels}

CHEOPSim is able to generate three type of bad pixel: dead, hot, and 'telegraphic'. Dead pixels are pixels with very low quantum efficiency. Hot pixels are pixels with anomalously high dark current. Telegraphic pixels {
(also known as random telegraph signal or RTS pixels)} are pixels which periodically flip between an active state with high dark current and an inactive state with normal dark current. The empirical dark frame contains seven hot pixels with dark current exceeding 5 electrons per second, of which one is located within the central 200$\times$200 sub-array. {
No dead or telegraphic pixels were observed during on-ground calibration. If bad pixels appear in the central region of the CCD during the mission, it is possible to choose a different target location on the CCD\footnote{
The target location is user configurable in CHEOPSim as described in Sect.~\ref{sec:fieldofview}.}.}

CHEOPSim can be configured to generate dead pixels with user defined sensitivities, and hot or telegraphic pixels with user defined dark current values, with locations assigned either manually, or randomly. For telegraphic pixels, the initial state is assigned randomly, and the period between one transition to the next (active to inactive, or inactive to active), is assigned assuming that the probability for a transition to occur after a given time is described by an exponential function~\citep{telegraphic}. Following each transition, the time until the next transition is drawn randomly from an exponential distribution with a user configurable time constant with a default value of nine minutes.

\subsubsection{Frame transfer smearing}

\begin{figure}[htbp]
\resizebox{\hsize}{!}{\includegraphics{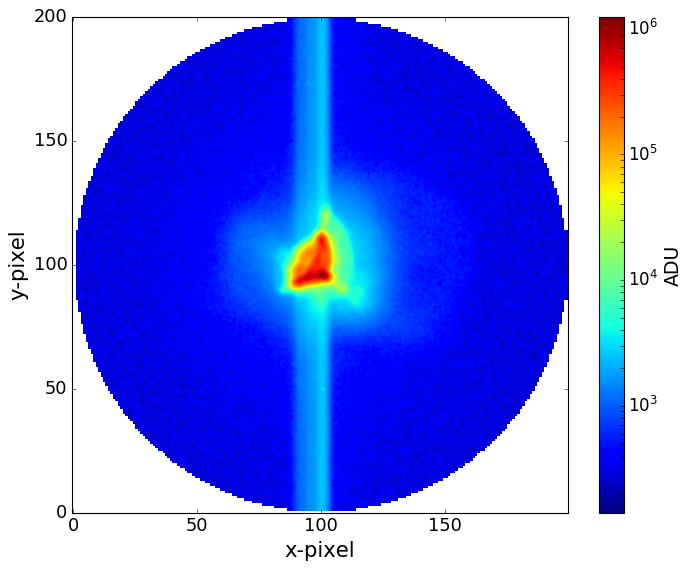}}
\caption{Sample image illustrating frame transfer smear trails, which are enhanced for very bright targets. In this example, the $V$-band magnitude of the target is 4.5, and the image is a stack of 39 exposures, each with duration 0.1s. The circular cut out to the 200$\times$200 pixel sub-array is applied in order to maximise science data down-link.}
\label{fig:frame_transfer}
\end{figure}

CHEOPS uses a frame transfer CCD. Frame transfer results in vertical smear trails being generated due to photons continuing to be incident on the CCD during the transfer. With the CCD oriented such that frame transfer occurs in a downward direction, there is a downward trail from the jittered PSF position at the start of the exposure and an upward trail from the jittered PSF position at the end of the exposure. The implementation for the trail in each direction is equivalent to making a sum over a series of images, each with exposure time equal to the frame transfer clock period (a configurable parameter with default value 25~$\mu$s), and each offset vertically by one pixel with respect to the previous image. Smear trails are similarly generated for hot and telegraphic pixels. The resulting smear trails are added to the main image. {
The content of overscan rows at the top of the CCD, which can be used in data processing to sample the frame transfer trails for each image, is also modelled.} Figure~\ref{fig:frame_transfer} shows an example image with visible smear trails.

\subsubsection{Shot noise}

Shot noise is the combination of two independent Poisson processes: photon noise arising from incident signal photons, plus shot noise arising from dark electrons. Since the processes are independent, the combined process is also a Poisson process, with expected value equal to the sum of the expected values of the individual processes. Thus, shot noise is applied to all pixels in the image by drawing randomly from a Poisson distribution with expected rate equal to the pixel value.

\subsubsection{Cosmic rays}
\label{sec:cosmics}

\citet{HST_ISR0207} reports that for the HST Advanced Camera for Surveys (ACS), between 1.5\% and 3\% of pixels are affected by cosmic rays in 1000 seconds. This implies an expected rate for CHEOPS of around ten cosmic ray events per minute in 200$\times$200 pixels of the central sub-array.

For each exposure in the simulation, the number of cosmic rays generated is drawn randomly from a Poisson distribution with expected rate $\lambda$ equal to a user configurable mean rate with default value 10.8 per minute in 200$\times$200 pixels, multiplied by the exposure time plus half the readout time in minutes. If the latitude and longitude of the spacecraft correspond to the location of the South Atlantic Anomaly, the expected rate is increased by a user specified factor with default value 1000.

For each cosmic ray, the following four parameters are generated randomly from a uniform distribution:
\begin {enumerate}
\item $x$-coordinate of impact position on the CCD surface
\item $y$-coordinate of impact position on the CCD surface
\item azimuthal angle of direction of travel with respect to CCD pixel coordinate system
\item polar angle of direction of travel with respect to CCD surface
\end{enumerate}

The length of the cosmic ray track projected onto the CCD surface is given by $d \tan\theta$, where $d$ is the thickness of the silicon substrate, equal to 15 microns and $\theta$ is the polar angle of the track. Starting from the impact position, the algorithm propagates along the length of the track in the direction defined by the azimuthal angle to determine the set of pixels through which the track passes, and the distance travelled through each pixel. Charge diffusion is not taken into account, meaning that charge is only deposited in pixels which are directly passed through by the cosmic ray.

The mean number of electrons produced per unit length is based on observations reported in~\citet{HST_ISR0207} for the HST ACS: Figures 3 and 4 in \citet{HST_ISR0207} show a sharp turn on in the distribution of the number of electrons at around 650 electrons, corresponding to cosmic rays with the minimum possible track length, which occurs for a normal angle of incidence, such that the length of the track is equal to the thickness of the silicon substrate. The expected number of electrons for a track with polar angle $\theta$ is thus taken to be $650/\cos\theta$. Given this expected value, the number of electrons in the simulated cosmic ray is generated randomly from a Landau distribution with location parameter $650/\cos\theta$ and scale parameter 150.  The scale parameter is chosen such that the resulting distribution for the number of electrons for cosmic rays in CHEOPSim is consistent with the corresponding distribution for the HST ACS High Resolution Channel CCD shown in Fig. 3 of \citet{HST_ISR0207}.

The total number of electrons for the track, generated as described above, is shared amongst the set of pixels through which the track passes according to the fraction of the total distance travelled within each pixel.  In order to account for Poisson fluctuations, the number of electrons assigned to each pixel is randomly smeared according to a Poisson distribution with expected value equal to the number of electrons before smearing.

\subsubsection{Full well saturation}
\label{sec:FullWellSimulator}

\begin{figure}[htbp]
\resizebox{\hsize}{!}{\includegraphics{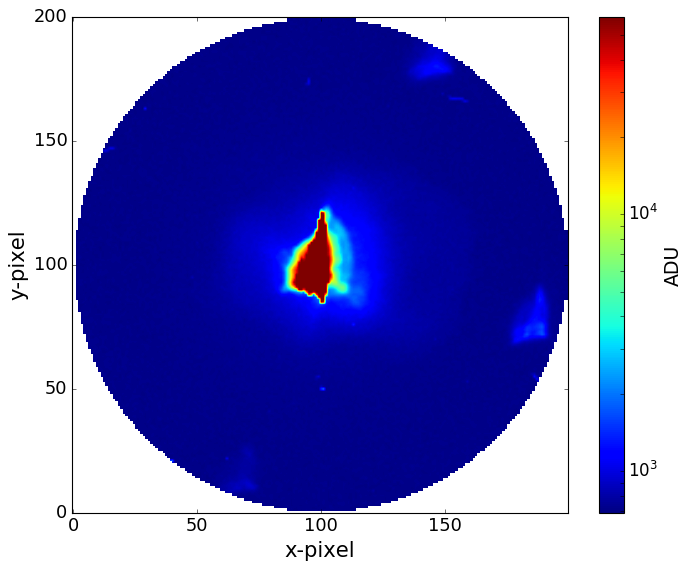}}
\caption{Example of a PSF with severe saturation and resultant vertical bleeding of the excess charges}
\label{fig:full_well_saturation}
\end{figure}

The CHEOPS CCD has a full well capacity of 121000 electrons\footnote{In practice the full well capacity is reduced to 114000 electrons in the central region of the CCD due to a 'peppering effect' that produces high frequency spatial variations when the flux reaches that level.}, defined as the number of electrons for which the deviation from linearity of the electronic gain (see Sect.~\ref{ccdNonLinearity}) reaches 3\%. Beyond this level, the pixel well begins to physically saturate, with electrons overflowing ('bleeding') into adjacent pixels. The number of electrons at which this occurs is not precisely known\footnote{Laboratory measurement was not possible due to the clamping effect described in sect.~\ref{sec:gain}.}. The threshold is configurable in CHEOPSim, with default value 125000. For pixels for which the number of electrons exceeds this value, the excess electrons bleed to adjacent pixels in the vertical direction, assuming an equal probability to bleed in either direction: half the excess electrons are shifted to the next pixel up, half are shifted to the next pixel down. If the adjacent pixel is also saturated, the charges continue to be moved until a pixel is reached which has remaining capacity, where the charges are deposited until it reaches saturation, and so on, until all the excess charges have been deposited. An example of a PSF with saturated pixels and resultant bleeding is shown in Fig.~\ref{fig:full_well_saturation}.

{
We note that in normal operations full well saturation should never occur for the target, because the exposure time is assigned separately for each target according to its magnitude and spectral type. It can occur for background stars in the field of view that are significantly brighter than the target.}

\subsubsection{Charge transfer efficiency}

Charge transfer efficiency (CTE) is the fraction of charges that are transfered from one pixel to the next during frame transfer and readout. At the start of the mission, the CTE is expected to be very high: 99.9999\% for vertical transfers and 99.9993\% for horizontal transfers. The process of shifting charges during frame transfer and readout, keeping track of charges left behind during each shift, is as follows:
\begin{enumerate}
\item Frame transfer: all the rows in the exposed part of the CCD are successively shifted down to the storage area in 1024 steps
\item Readout: the rows in the storage area are successively shifted down, and after each shift down, the pixels in the bottom row (readout register) are successively shifted to the left. After each shift left, the value of the leftmost pixel of the bottom row is used to assign the value of the pixel (i,j) in the image after readout, where i and j are the number of horizontal and vertical shifts, respectively, that have so far taken place during the frame transfer and readout.
\end{enumerate}

For each pixel shift in the procedure outlined above, if $i$ is the index of the pixel to be shifted and $i-1$ is the index of the destination pixel, the number of electrons, $N_\mathrm{e}[i-1]$ assigned to the destination pixel is given by:
\begin{equation}
N_\mathrm{e}[i-1] = \mathrm{CTE}\times N_\mathrm{e}[i] + (1-\mathrm{CTE})\times N_\mathrm{e}[i-1].
\end{equation}

\subsubsection{Charge transfer efficiency at end of life}

Over time, cosmic ray hits result in damage to the silicon lattice, giving rise to 'charge traps': charges left behind during a transfer are not directly available for the next transfer as in the beginning of life case but are trapped and released according to an exponential decay with a time constant which is independent of the number of charges left behind. As a result, point sources will show up with extended tails. The fraction of the pixel charge which is transferred to the tail is signal dependent since the charge traps have a capacity to hold a greater fraction of a small charge. Tails generated during the relatively fast frame transfer dominate over the position dependent tails generated during the slower shifting of pixels during readout. The model described here takes into account frame transfer only.

The model\footnote{The model used is very simple, and by no means describes the full complexity of real CTI effects resulting from charge trapping. For example, each pixel is considered independently from its neighbours, whereas in reality there is a strong interaction due to traps being filled by charges from pixels in preceding rows. Also background is not considered, the presence of which results in partial filling of the traps, resulting in reduced smearing of the signal.} is based on properties observed from proton irradiated detectors in laboratory experiments~\citep{Verhoeve}, in which the signal dependence of the charge transfer inefficiency, $\mathrm{CTI}=1-\mathrm{CTE}$, was observed to follow a power law:
\begin{equation}
\mathrm{CTI}(S) = \mathrm{CTI_0}(S/S_0)^K,
\end{equation}
where $S$ is the number of charges in the pixel and $\mathrm{CTI_0}$ is a constant corresponding to the fraction of charges that will be transferred to the tail for $S=S_0$. Based on measurements on the PLATO CCD, default values for the configurable constants $K=-0.65$ and $\mathrm{CTI_0}=0.028$ for $S_0=10000$ are derived.

For a given pixel, the fraction of the charge in the pixel that will be transferred to the tail is determined according to the power law. The CTI fraction is subtracted from the original pixel and distributed in the pixels above it according to an exponential distribution with a configurable decay constant, with default value 100 pixels.

\begin{figure}[htbp]
\resizebox{\hsize}{!}{\includegraphics{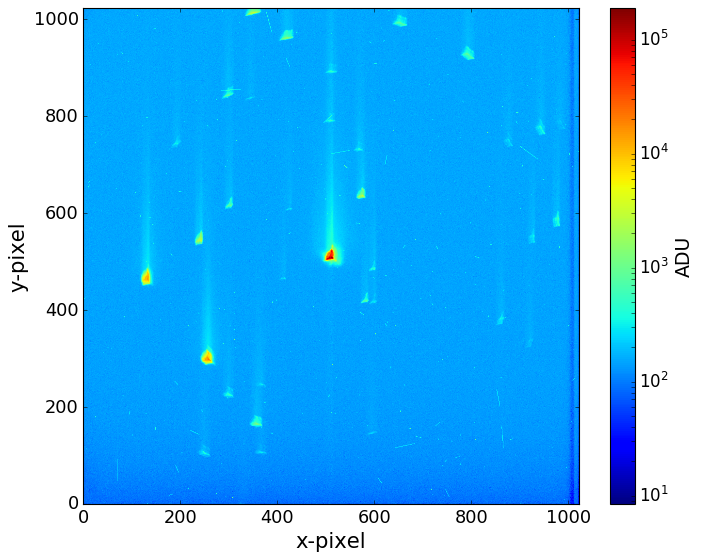}}
\caption{Effect of the end of life CTI for a stack of nine exposures with exposure duration 4 seconds, for a 9th magnitude target (in the centre). For comparison, the frame transfer smear trail is just visible below the target star.}
\label{fig:cti_eol}
\end{figure}

The effect of the end of life CTI is shown in Fig.~\ref{fig:cti_eol}.

\subsubsection{CCD non-linearity}
\label{ccdNonLinearity}

Measurements performed during on-ground calibration have been used to derive a correction to account for non-linearity of the CCD response in the form of a quadratic spline \citep{Deline}. Measurements were performed separately for readout rates 230~kHz and 100~kHz. CCD non-linearity in CHEOPSim is modelled by applying the inverse of this correction. The spline function is composed of ten intervals for 230~kHz and six intervals for 100~kHz, and is defined by the coefficients of a second order polynomial within each interval. The boundaries between the intervals are defined in terms of uncorrected numbers of electrons. The input to the inverse correction in CHEOPSim is corrected numbers of electrons, so the interval boundaries are mapped to corrected numbers of electrons for each readout rate. For a corrected pixel value within a given corrected number of electrons interval, the uncorrected value is calculated by solving the quadratic equation for the interval concerned.

\begin{figure}[hbtp]
\resizebox{\hsize}{!}{\includegraphics{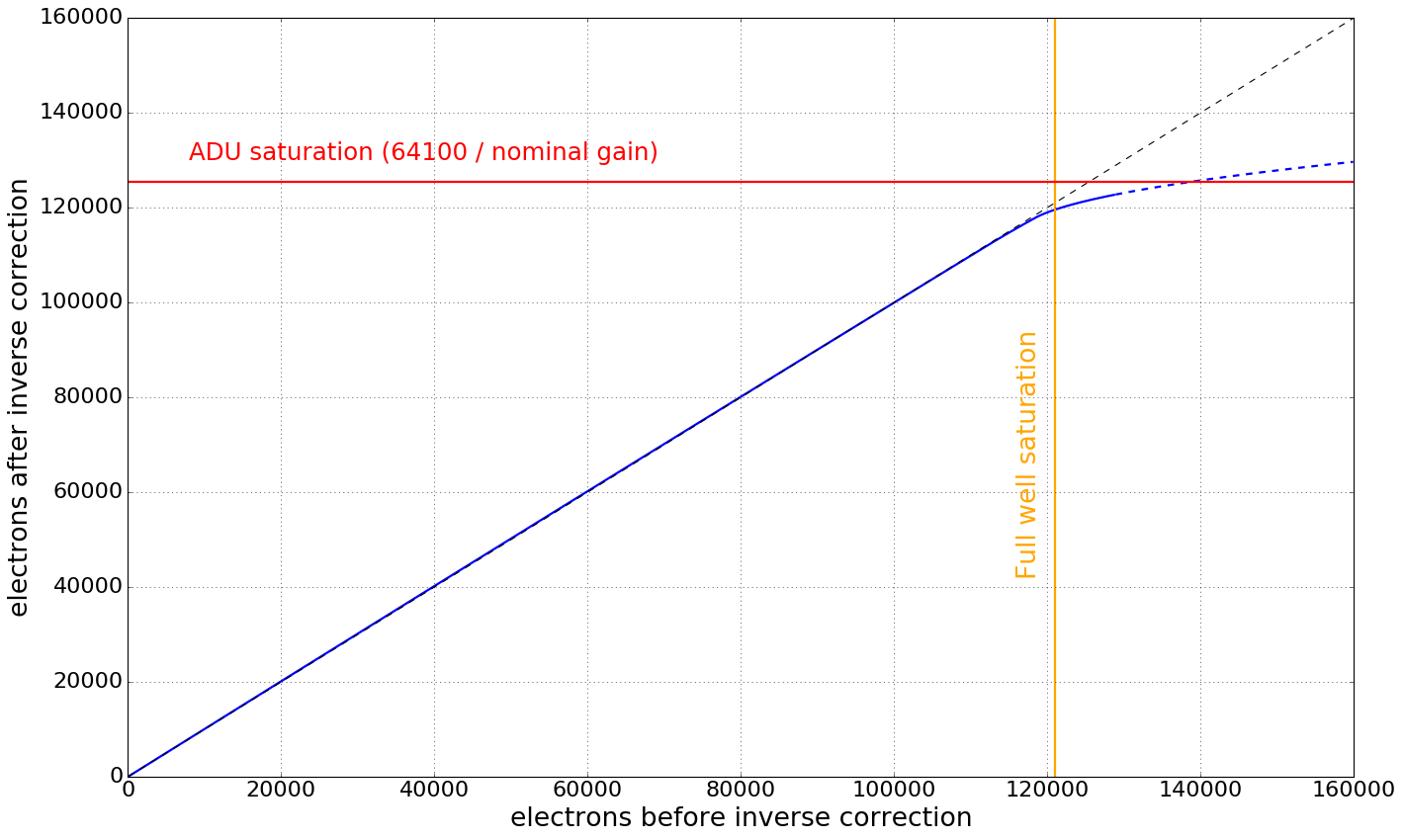}}
\caption{Inverse correction function used to define the CCD non-linearity for 100~kHz readout rate. The inverse correction function is extended to higher number of electrons than the last interval of the spline function, by extending the polynomial defined for the last interval (dashed blue line).}
\label{fig:ccdNonLinearity}
\end{figure}

The inverse correction function is shown for readout rate 100~kHz in Fig.~\ref{fig:ccdNonLinearity}. It shows the number of electrons after applying the inverse correction (non-linearity applied) as a function of the number of electrons before applying the inverse correction (non-linearity not applied). The inverse correction function is extended to higher numbers of electrons than the last interval of the spline function, by extending the polynomial defined for the last interval (dashed blue line).

\subsubsection{Electronic gain, bias offset and read noise}
\label{sec:gain}

The units of the pixel values in the image are converted from electrons to analog-to-digital units (ADU) by multiplying by the electronic gain. The nominal gain of the CHEOPS CCD is 0.5111 ADU per electron. The gain has a dependence on four bias voltages and on the CCD temperature. During ground calibration, the four voltages have been observed to drift as a function of time. This drift is modelled in CHEOPSim, resulting in a drift of the gain value with time. In order to model measurement uncertainty on the voltages, and therefore uncertainty on the gain, the values appearing in the image metadata and in the housekeeping data are additionally subjected to random Gaussian smearing, with Gaussian widths corresponding to the RMS observed during the payload calibration. The CCD temperature is assumed to have no drift, but Gaussian fluctuations are applied in the image metadata and housekeeping data.

The final step of the simulation is to apply a bias level offset and readout noise to the image by adding to each pixel a value obtained by drawing randomly from a Gaussian distribution, whose mean (corresponding to the bias offset) and width (corresponding to the readout noise) are assigned separately for each pixel according to an empirical bias frame obtained during the on-ground calibration \citep{Deline}.

The maximum possible ADU count for each pixel in an unstacked image is $2^{16}-1=65535$ for the 16 bit ADC. Pixels close to ADC saturation, defined as having ADU count above 64100, are clamped to the maximum value of 65336 in order to avoid unexpected ADC saturation and to flag unexpected pixel behavior. A drawback of this feature is that the charge information is lost when a pixel approaches its saturation limit, which prevents saturation photometry. {
We note, however, that saturation photometry is not needed for CHEOPS because the exposure time is assigned separately for each observation of each target according to their magnitude and spectral type, such that it is always ensured that saturation does not occur for the target.}


\section{Application: Photometric extraction for simulated exoplanet transits}

\subsection{Validation of CHEOPS photometric extraction performance using simulated targets}
\label{sec:light_curves_SV3}

A validation campaign has been carried out for the data reduction software \citep{Hoyer}, which will be used to calibrate and correct images and to extract light curves from image time series for real data. The validation used simulated image time series generated by CHEOPSim, with the aim of demonstrating that the scientific requirements for the mission are satisfied.

\subsubsection{Transit detection}
\label{sec:req1}
The {
science requirements specify} that CHEOPS shall be able to detect Earth-sized planets transiting G5 dwarf stars with $V$-band magnitudes in the range $6\leq V\leq9$ mag. Since the depth of such
transits is 100 parts-per-million (ppm), this requires achieving a photometric precision of
20 ppm in six hours of integration time. This time corresponds to the transit
duration of a planet with a revolution period of 50 days. The primary targets of this requirement are stars that,
having radial velocity measurements, are already known to host Earth-sized planets. This requirement guarantees a signal-to-noise ratio of 5, which allows planet transits to be reliably detected.

\begin{figure}[hbtp]
\resizebox{\hsize}{!}{\includegraphics{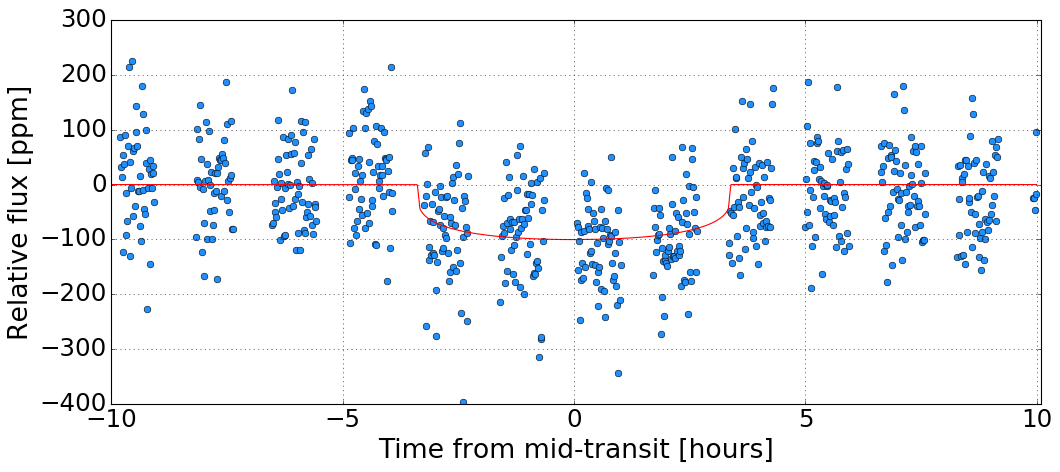}}
\resizebox{\hsize}{!}{\includegraphics{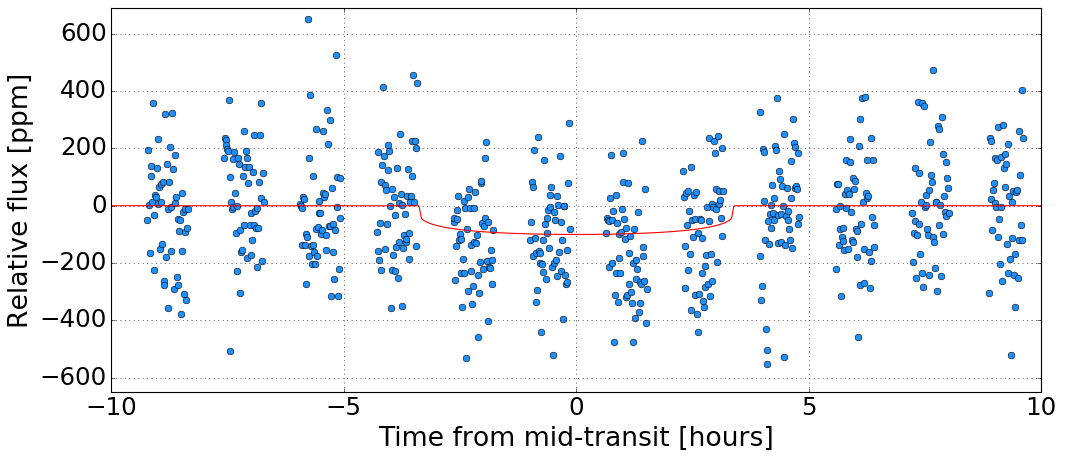}}
\caption{Light curves extracted from CHEOPSim image time series, using photometric
extraction performed by the data reduction pipeline that will be used with real CHEOPS data, for 
an Earth-sized planet with a 50 day orbital period orbiting a $V$=6 G2V star (top), and for the same planet orbiting a $V$=9 G2V star (bottom). For the $V$=6 case, the exposure duration is 0.5 seconds, and exposures are stacked with 40 exposures per stack. For the $V$=9 case, the exposure duration is 10 seconds, and exposures are stacked with six exposures per stack.
Gaps in the light curves correspond to interruptions due to Earth occultation (once per CHEOPS orbit). The simulated photon flux incident on the telescope is shown in each case by the red line. Stellar granulation was not simulated, resulting in smooth curves.}
\label{fig:transit_detection}
\end{figure}

Two datasets were generated in order to test this requirement, each corresponding to a 20 hour visit for a Sun-like target (spectral type G2V), with magnitudes $V$=6 and $V$=9, respectively. Both simulations included the transit of an Earth-sized planet with an orbital period of 50 days. The exposure durations were 0.5 seconds and 10 seconds, respectively. The light curves extracted using the default 33 pixel aperture radius are shown in Fig.~\ref{fig:transit_detection}.

\subsubsection{Transit characterisation}
\label{sec:req2}
The {
science requirements specify} that CHEOPS shall be able to detect Neptune-sized planets transiting K-type dwarf stars with $V$-band magnitudes as faint as $V$=12 mag with a
signal-to-noise ratio of 30. Such transits have depths of 2500 ppm and last for nearly
three hours, for planets with a revolution period of 13 days. Hence, a photometric precision
of 85 ppm is to be obtained in three hours of integration time.
The primary targets of this requirement are hot and warm Neptunes already known to transit their parent star.
This requirement guarantees a signal-to-noise ratio of 30, which will enable a detailed characterisation\footnote{
The precise measurement of the planet radius, where the mass is known from ground based observations, provides characterisation of the planets' internal structure through the determination of the bulk density.} of the transit light curve.

\begin{figure}[hbtp]
\resizebox{\hsize}{!}{\includegraphics{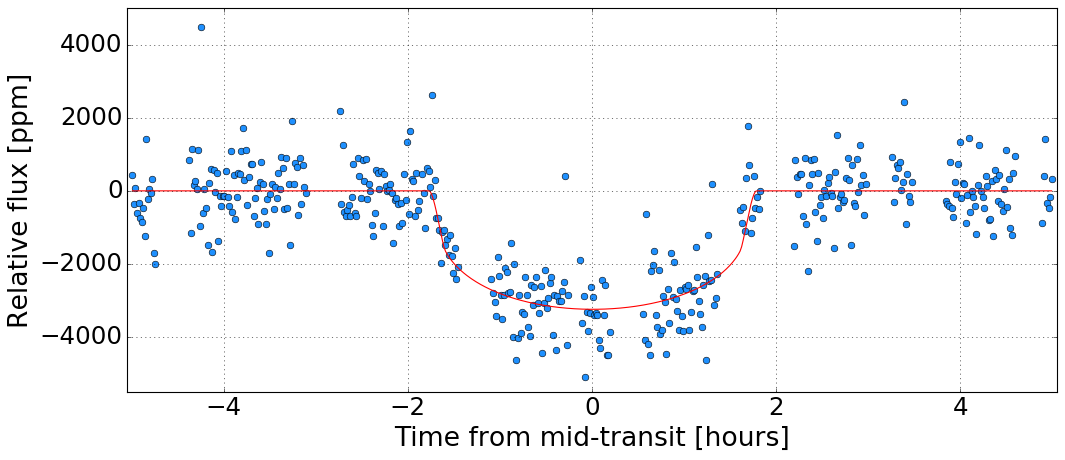}}
\caption{Light curve extracted from a CHEOPSim image time series, using photometric
extraction performed by the data reduction pipeline that will be used with real CHEOPS data, for a Neptune-sized planet orbiting a $V$=12 K5V star with a 13 day orbital period. The exposure duration is 60 seconds and exposures are not stacked. Gaps in the light curve correspond to interruptions due to Earth occultation (once per CHEOPS orbit). The simulated photon flux incident on the telescope is shown by the red line. Stellar granulation was not simulated, resulting in a smooth curve.}
\label{fig:transit_characterisation}
\end{figure}

A dataset was generated corresponding to a ten hour visit for a magnitude $V$=12 target with spectral type K5V, orbited by a Neptune-sized planet with an orbital period of 13 days. The exposure duration was 60 seconds. The light curve extracted using the default 33 pixel aperture radius is shown in Fig.~\ref{fig:transit_characterisation}.

\subsubsection{End-to-end validation of the data processing chain}

For each of the datasets in Sects.~\ref{sec:req1} and \ref{sec:req2}, four independent analysts from the CHEOPS Science Team were asked to apply their fitting analysis to extract the planet radii from the light curves. 

The purpose of this validation exercise was to simulate the data processing chain from end to end. The data products generated by CHEOPSim were first processed with the on-board software to convert to the binary format in which the data is set to ground. The raw data was then processed through the on-ground processing chain, ending with the data-reduction pipeline, and finally ingested into the data archive. The analysts tested the archive functionality, and were instructed to analyse the data using their preferred data-processing routines. The only artificial constraints imposed were that the impact parameter of the transits should be fixed at zero, and that the photometric aperture should be fixed at 33 pixels for uniformity of background treatment for science-requirements compliance testing.

Different models and parameter estimation methods were used by each analyst. The transit models were the Mandel-Agol method \citep{mandelxx}, {\sc batman} \citep{kreidbergxx} and {\sc qpower2} \citep{maxtedxx1}, with either quadratic limb darkening or the power-2 formulation \citep{maxtedxx2}. Linear or quadratic decorrelation of the image centroid position on the CCD background level was carried out in some analyses, which showed weak linear dependences on the image centroid position and a weak quadratic dependence on the background level. Parameter estimation was carried out with {\sc emcee} \citep{foremanmackeyxx}, {\em MC3} \citep{cubillosxx} or custom MCMC routines implemented by the analysts. Overall, the analyses represented a sample of the likely data-analysis preferences of the CHEOPS user community. The procedure therefore tested compliance at the end of a data-processing chain that included the variety of data-analysis approaches expected from the user community. 

The median measured values of the planet-to-star radius ratio are compared to the true values in Table~\ref{tab:planet_radii}. The measured values are consistent with the true values to within two standard deviations. The precision and accuracy are compliant with the mission requirements.

\begin{table*}
\caption{Planet-to-star radius ratios measured from the light curves shown in Figs.~\ref{fig:transit_detection} and \ref{fig:transit_characterisation}, compared to the true values that were input to the simulations. The measured values correspond to the median of the measurements from four independent analyses. The statistical uncertainty corresponds to the mean of the statistical uncertainties for the four analyses. The systematic uncertainty corresponds to the standard deviation of the measurements for the four analyses. {
We note that since the same flux time series was used for all analyses, the statistical and systematic uncertainties as they are defined are not independent, so cannot be added in quadrature to obtain a total uncertainty. The conservative approach of adding the two errors yields a total error which still meets the requirements.}}
\label{tab:planet_radii}
\centering
\begin{tabular}{c c c c c}
\hline\hline
Target $V$-mag & Planet size & No. of transits & Measured $R_p/R_s$ & True $R_p/R_s$ \\
\hline
6 & Earth & 1 & 0.00923 $\pm$ 0.00054(stat) $\pm$ 0.00019(syst) & 0.00916 \\
9 & Earth & 2 &  0.01012 $\pm$ 0.00068(stat) $\pm$ 0.00024(syst) & 0.00916 \\
12 & Neptune & 1 & 0.05038 $\pm$ 0.00061(stat) $\pm$ 0.00031(syst) & 0.05000 \\
\hline
\end{tabular}
\end{table*}


\subsection{Photometric extraction performance for real targets}

In this Section we present some additional examples of light curves extracted from image time series generated with CHEOPSim, in order to further demonstrate the potential capabilities of CHEOPS for exoplanet detection in some more challenging scenarios, including cases outside the mission requirements. As for Sect.~\ref{sec:light_curves_SV3}, photometric extraction is performed using the same data reduction software \citep{Hoyer} as will be used for real data.

\subsubsection{Transit for a star with a nearby companion}

An interesting target for CHEOPS is the planet HD 80606b, a Jupiter-sized planet (0.987$R_\mathrm{J}$) which has a highly elliptical orbit (eccentricity 0.9336) around its parent star, which is part of a binary system composed of the stars HD 80606 and HD 80607. The two stars have near identical magnitudes: $V$=9.00 and $V$=9.01, respectively, and both are spectral type G5V. The orbital period of the planet is 111 days, with a transit due in February 2020, making this a potential early target for CHEOPS.

CHEOPSim was used to generate a simulated image time series for an observation of the system. The light curve extracted using the default 33 pixel aperture radius is shown by the dark blue points in Fig.~\ref{fig:HD80606}. The observed transit depth is half the theoretical depth due to the diluting effect of the binary companion star being within the photometric aperture. A sawtooth pattern is also visible in the extracted light curve, which is a result of variation with field of view rotation of the fraction of the irregularly shaped PSF of the companion star which lies within the circular photometric aperture\footnote{The orientation of the PSF does not rotate as the field of view rotates.}. Despite these (correctable) effects, the transit is clearly observable. The sawtooth pattern can be eliminated, without significantly increasing the overall noise in the light curve, by using an aperture radius large enough to fully contain the flux from the companion (pale blue-grey points in Fig.~\ref{fig:HD80606}). It is possible that for a case such as this, the centroiding may confuse the companion for the target, putting the companion at the centre of the image for part (or all) of the observation. However, this does not matter if an aperture large enough to contain both stars is used. \citet{Hoyer} describe a procedure to determine an optimal aperture radius for which the signal-to-noise is maximised, and indeed for the case of this target, the optimised radius is large enough to fully contain the flux from both stars.

\subsubsection{Transit for a very faint target}

CHEOPSim has been used to demonstrate the potential capability of CHEOPS to detect super-Earths around stars significantly fainter than {
the magnitudes specified in the science requirements}. Figure~\ref{fig:GJ1214} shows the extracted light curve for the planet GJ~1214b, which has radius 2.68$R_\Earth$, and an orbit period of 1.58 days around its parent star, which has magnitude $V$=15.1 and spectral type M4.5V. The clearly visible dip in the extracted light curve shows that there is potential for such planets to be detectable by CHEOPS. We note, however, that for such faint targets, it is important that there are no brighter stars within 30-40 arcseconds of the target, otherwise the target acquisition and centroiding are likely to fail.

\subsubsection{Transit for a very bright target}

Finally, we present an example of an extracted light curve for the case of a target star with magnitude brighter than {
specified in the CHEOPS science requirements}. The target chosen for this case is the star Pi Mensae. It is important to note that in reality this target cannot be observed by CHEOPS due to its location close to the Southern Ecliptic Pole, a region which is excluded due to the stringent 120\textdegree\ Sun-avoidance exclusion angle. The target has nonetheless been chosen as it allows comparison of the photometric abilities of the CHEOPS and TESS missions. Stray light from the Sun is not modelled in the simulation. Figure~\ref{fig:PiMen} shows the extracted light curve for the planet $\pi$ Men c, which has radius 2.04$R_\Earth$, and an orbit period of 6.27 days around its parent star Pi Mensae, which has magnitude $V$=5.67 and spectral type G0V. For comparison, Fig.~\ref{fig:PiMen} also shows the phase folded light curve obtained during the discovery of the planet by TESS  \mbox{\citep{PiMen}}. It can be seen that for a single transit, the expected photometric precision for CHEOPS exceeds that of TESS.

\begin{figure}[hbtp]
\resizebox{\hsize}{!}{\includegraphics{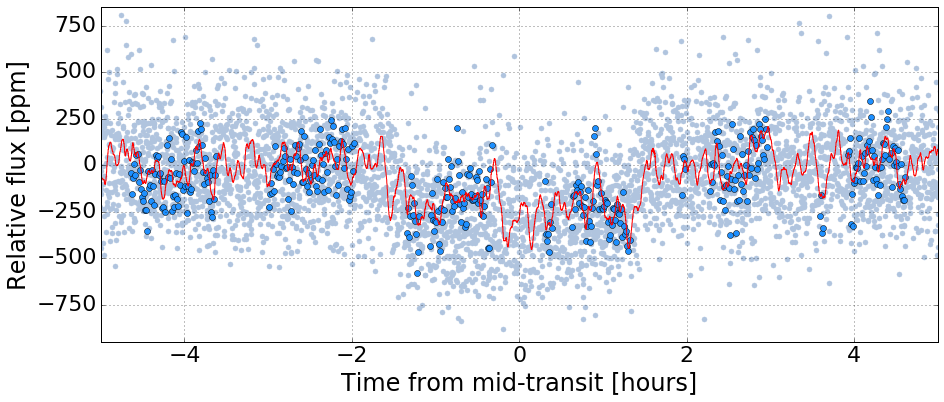}}
\caption{Dark blue points: light curve extracted from a CHEOPSim image time series, using photometric
extraction performed by the data reduction pipeline that will be used with real CHEOPS data, for planet $\pi$ Men c (radius 2.04$R_\Earth$, orbital period 6.27 days), orbiting the star Pi Mensae ($V$=5.67, spectral type G0V). The exposure duration is 0.35 seconds and the exposures are stacked with stacking order 33. Gaps in the light curve correspond to interruptions due to Earth occultation (once per CHEOPS orbit). The simulated photon flux from the target star incident on the telescope, with stellar granulation as the only noise source, is shown by the red line. {
The pale blue-grey points show the phase folded light curve for the planet $\pi$ Men c as measured by TESS \citep{PiMen}.}}
\label{fig:PiMen_withGranulation}
\end{figure}

\begin{figure*}[hbtp]
\resizebox{\hsize}{!}{\includegraphics{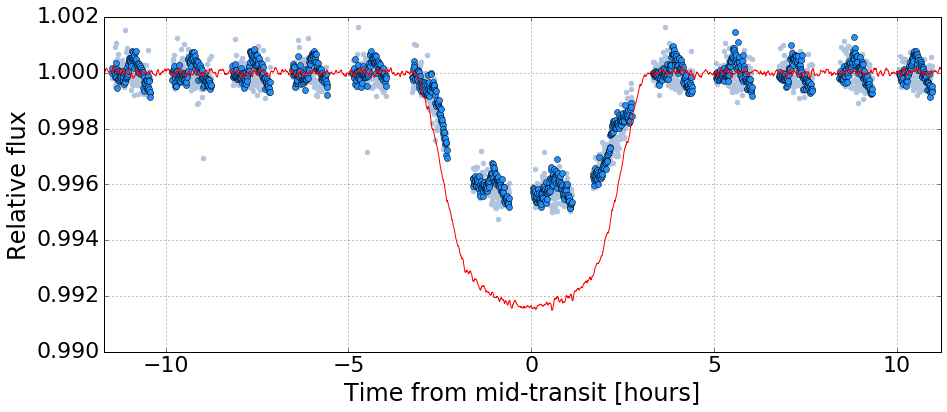}\includegraphics{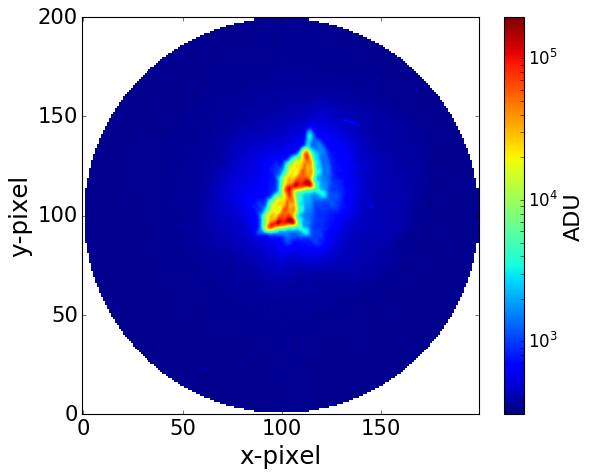}}
\caption{Light curve extracted from a CHEOPSim image time series, using photometric
extraction performed by the data reduction pipeline that will be used with real CHEOPS data, for planet HD 80606b (radius 0.987$R_\mathrm{J}$, orbital period 111 days), orbiting the star HD80606 ($V$=9.00, spectral type G5V). The exposure duration is 7 seconds and the exposures are stacked with stacking order 5. Gaps in the light curve correspond to interruptions due to Earth occultation (once per CHEOPS orbit). The simulated photon flux from the target star incident on the telescope, with stellar granulation as the only noise source, is shown by the red line. The transit depth from the photometric extraction is diluted by factor 2 due to the presence of the binary companion star HD 80607 within the photometric aperture (radius 33 pixels), visible in the example image shown on the right. The sawtooth pattern is due to variation of the fraction of the irregularly shaped PSF of the companion star which lies within the circular photometric aperture as the field of view rotates. The pale blue-grey points show the light curve from the same image data, using a larger photometric aperture radius of 92.3 pixels, such that the all the flux from both stars is fully contained. In this case, the sawtooth pattern is absent.}
\label{fig:HD80606}

\resizebox{\hsize}{!}{\includegraphics{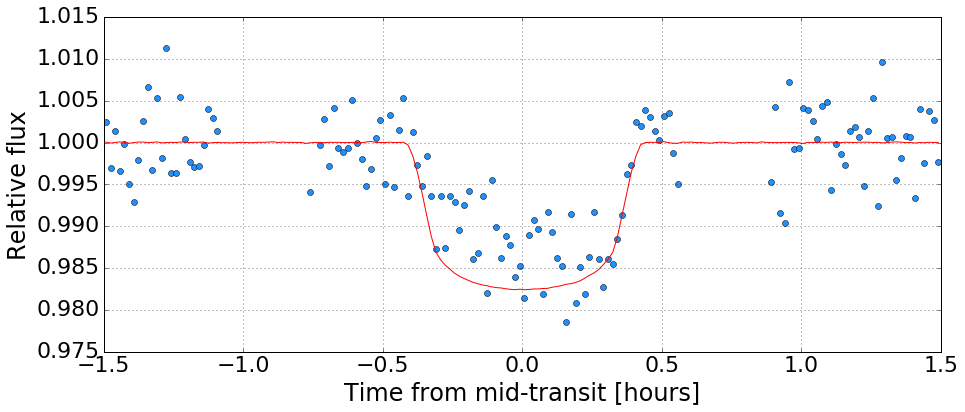}\includegraphics{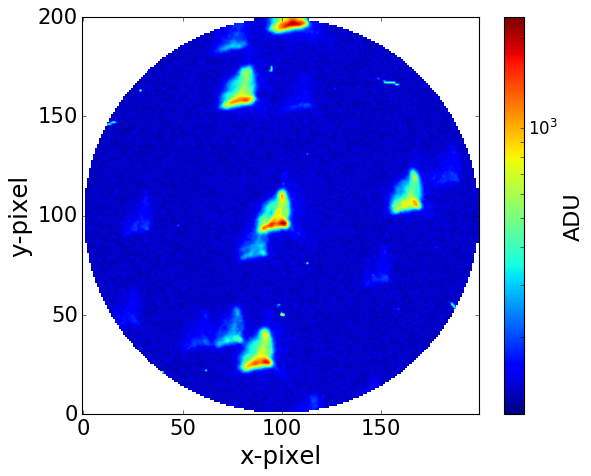}}
\caption{Light curve extracted from a CHEOPSim image time series, using photometric
extraction performed by the data reduction pipeline that will be used with real CHEOPS data, for planet GJ 1214b (radius 2.68$R_\Earth$, orbital period 1.58 days), orbiting the star GJ 1214 ($V$=15.1, spectral type M4.5V). The exposure duration is 60 seconds and exposures are not stacked. Gaps in the light curve correspond to interruptions due to Earth occultation (once per CHEOPS orbit). The simulated photon flux from the target star incident on the telescope, with stellar granulation as the only noise source, is shown by the red line. The transit depth from the photometric extraction is diluted due to the presence of a contaminant background star within the photometric aperture (radius 33 pixels), visible in the example image shown on the right.}
\label{fig:GJ1214}

\resizebox{\hsize}{!}{\includegraphics{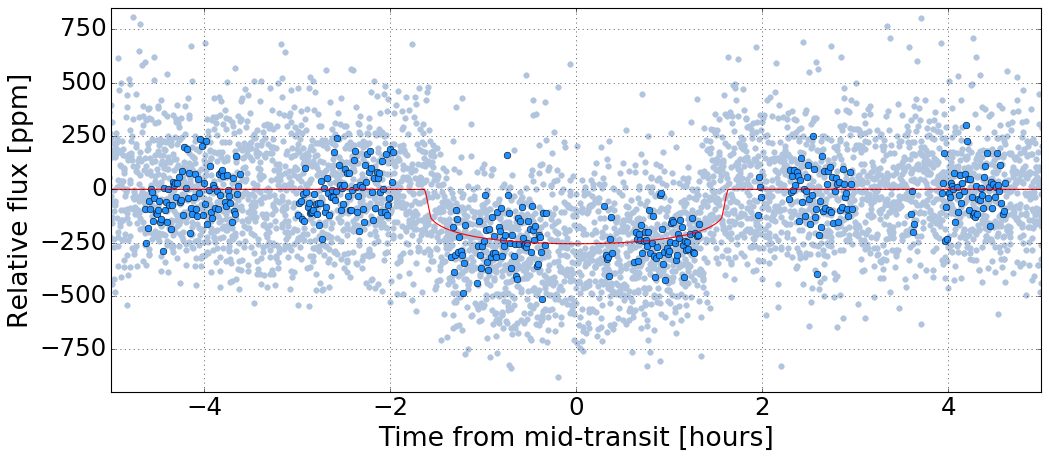}\includegraphics{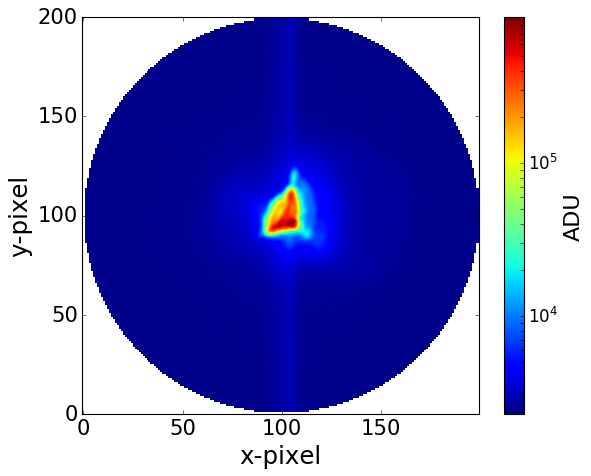}}
\caption{Dark blue points: light curve extracted from a CHEOPSim image time series, using photometric
extraction performed by the data reduction pipeline that will be used with real CHEOPS data, for planet $\pi$ Men c (radius 2.04$R_\Earth$, orbital period 6.27 days), orbiting the star Pi Mensae ($V$=5.67, spectral type G0V). The exposure duration is 0.35 seconds and the exposures are stacked with stacking order 33. In reality this target cannot be observed by CHEOPS due to its location close to the Southern Ecliptic Pole, a region which is excluded due to the 120\textdegree\ Sun-avoidance exclusion angle, but has been chosen as it allows comparison of the photometric abilities of the CHEOPS and TESS missions. Stray light from the Sun has not been modelled. Gaps in the light curve correspond to interruptions due to Earth occultation (once per CHEOPS orbit). The gaps here are close to worst case, since the target is close to the Southern Ecliptic Pole, and CHEOPS has a polar Earth orbit. The simulated photon flux from the target star incident on the telescope, without stellar granulation, is shown by the red line. An example image is shown on the right. Pale blue-grey points: phase folded light curve for the planet $\pi$ Men c as measured by TESS \citep{PiMen}. }
\label{fig:PiMen}
\end{figure*}

Unlike the simulations for HD80606 and GJ1214, the $\pi$ Men simulation does not include stellar granulation (Sect.~\ref{sec:stellar_variation}). This choice was made in order to simplify the comparison with the TESS result, since the true level of stellar granulation for the star is not known. The same simulation was also run with stellar granulation included, and the result is shown in Fig.~\ref{fig:PiMen_withGranulation}. Comparing the extracted light curve with the simulated incident photon flux shows that for a target this bright, the achievable precision is limited by stellar granulation.

\section{Conclusions}

In this paper, we present the CHEOPS simulator used to generate simulations of the data which will be received from the CHEOPS satellite. We present descriptions of the methods used for detailed modelling of the incident flux, the satellite orbit and pointing jitter, the telescope optics, and the response of the CCD.

We present results from the use of CHEOPSim data to validate the data reduction processing chain, which has been used to generate light curves from CHEOPSim data with simulated planetary transits. Independent analysts were successfully able to detect the planets from the data in these light curves and measure their radii, using a variety of models and parameter estimation methods, to an accuracy within the mission science requirements. 

We also used CHEOPSim to explore the range of capabilities of CHEOPS beyond the mission requirements by simulating some real targets with known planets. We found that a deep transit for a star fainter than requirements could be detectable by CHEOPS and that the photometric performance for a star brighter than requirements exceeds that of TESS, and it is limited by stellar granulation.

These expected performances, and the performance of the spacecraft more generally, will be verified during the in flight commissioning.

\begin{appendix}
\section{Output}
\label{app:output}

According to the configuration, the output of CHEOPSim includes the following:
\begin{enumerate}
\item A set of FITS files corresponding to the output of the pre-processing of RAW data at the Software Operations Centre (SOC), which can be used as input to the Quick Look software for fast inspection of the data, and to the data reduction processing chain. This consists of: a full frame exposure at the start of the visit; a time series of stacked images in the form of a FITS image cube, including the content of the CCD margins (dark, blank and overscan reference columns/rows) as separate FITS extensions; a time series of imagettes (circular cut outs with default diameter 35 pixels) with the cadence of individual exposures; a FITS table containing a time series of centroids with the cadence of the exposures; FITS files containing orbit and attitude information; and FITS files containing housekeeping data.\label{item1}
\item A set of files providing a time series of unstacked images as FITS image cubes which can be used as input to the Data Flow Simulator, which which executes the on-board software to compress the images into the bitstream format that will be sent to ground from the spacecraft, thus providing simulated data to test
the pre-processing of the RAW data at the SOC in the form in which it will be received from the Mission Operations Centre (MOC). In testing this procedure, it is required as a closure test that the output of the pre-processing chain matches the output described in~\ref{item1}.
\item FITS tables containing the incident photon flux for each exposure, with and without photon noise applied. The time series with photon noise represents the ideal light curve which could be extracted from the images if detector effects could be perfectly corrected, and thus represents a reference for evaluating the performance of photometric extraction.
\item FITS tables providing truth information about the simulated data, including the generated positions of PSFs and bad pixels, the incident flux and contributions from background sources, and the roll angle, together with image cubes containing information for cosmic rays and smear trails prior to the application of noise.
\end{enumerate}

\section{Configuration and execution}
\label{app:web}

\begin{figure}[hbtp]
\resizebox{\hsize}{!}{\includegraphics{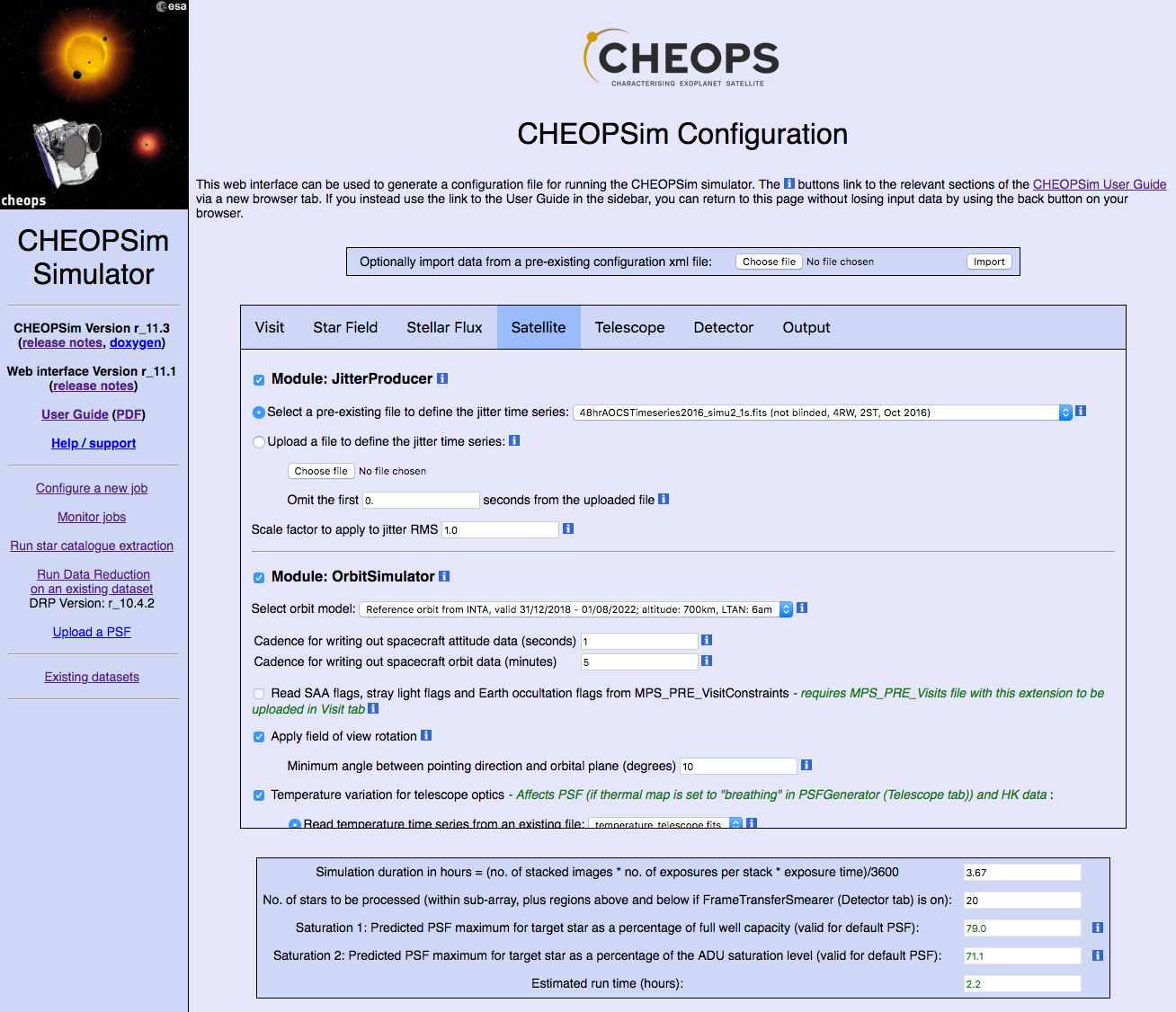}}
\caption{Screen shot of the web interface used to configure CHEOPSim. The interface is divided into tabs, of which part of the Satellite tab is displayed, for configuring the parameters relating to jitter and orbit.}
\label{fig:web}
\end{figure}

The configuration of CHEOPSim is defined using an xml file with over one hundred configurable parameters. The xml file is generated by the user via a web interface.
A screenshot of the web interface is shown in Fig.~\ref{fig:web}.
As parameters are adjusted, the web interface dynamically displays the number of stars in the field of view, the predicted saturation level for the target star, and an estimate of the processing time. Once the configuration file has been generated, a preview of the field of view is displayed.

The web interface provides a tool to extract a list of stars within the field of view from the Gaia star catalogue, which can then be used as input to the simulation. It also provides an interface to the mission planning software, which determines the scheduling sequence for observations and defines each observation via a set of parameters: the output from the mission planning for a given observation can be uploaded to the form and used to automatically assign CHEOPSim parameter values. The interface also provides the option to run the data reduction processing chain on the output, in order to extract a light curve.

Submitted configurations are executed on a computing cluster at the University of Geneva, and thus no software installation by the user is required. The status of submitted jobs is tracked using a PostgreSQL database, and live progress can be monitored via the web interface. Upon job completion, the user receives an email indicating the location of the job output on a public ftp server.

\end{appendix}

\begin{acknowledgements}
We are very grateful the following people for their input to the development of CHEOPSim: Roi Alonso, Rodrigo Diaz, Helen Giles, Claudia Greco, Kate Isaak, Demetrio Magrin, Göran Olofsson, Mahmoudreza Oshagh. {
We thank ESA (Carlos Corral Van Damme and Andrew Hyslop) for providing jitter time series and for assistance in the implementation and validation of the orbit model. We also thank ESA (in particular Peter Verhoeve) for the quantum efficiency measurements and INAF and Leonardo S.p.A. for the optical throughput measurements.} We also thank the users of CHEOPSim for their helpful feedback. This project has received funding from the European Research Council (ERC) under the European Union’s Horizon 2020 research and innovation programme (project {\sc Four Aces}; grant agreement No 724427). It has also been carried out in the frame of the National Centre for Competence in Research PlanetS supported by the Swiss National Science Foundation (SNSF). {
We also acknowledge support from the Swiss Space Office (SSO).}
\end{acknowledgements}

\bibliographystyle{aa} 
\bibliography{cheopsim_paper.bib} 

\end{document}